# Practical relevance of software engineering research: Synthesizing the community's voice

| Vahid Garousi | Markus Borg | Markku Oivo |
|---|---|---|
| Queen's University Belfast | RISE Research Institutes of | M3S group |
| Northern Ireland, UK | Sweden AB, Sweden | University of Oulu, Finland |
| v.garousi@qub.ac.uk | markus.borg@ri.se | markku.oivo@oulu.fi |

**Abstract:**

Software engineering (SE) research should be relevant to industrial practice. There have been regular discussions in the SE community on this issue since the 1980's, led by pioneers such as Robert Glass. As we recently passed the milestone of "50 years of software engineering", some recent positive efforts have been made in this direction, e.g., establishing "industrial" tracks in several SE conferences. However, many researchers and practitioners believe that we, as a community, are still struggling with research relevance and utility. The goal of this paper is to synthesize the evidence and experience-based opinions shared on this topic so far in the SE community, and to encourage the community to further reflect and act on the research relevance. For this purpose, we have conducted a Multi-vocal Literature Review (MLR) of 54 systematically-selected sources (papers and non peer-reviewed articles). Instead of relying on and considering the individual opinions on research relevance, mentioned in each of the sources, the MLR aims to synthesize and provide the "holistic" view on the topic. The highlights of our MLR findings are as follows. The top three root causes of low relevance, discussed in the community, are: (1) Researchers having simplistic views (or wrong assumptions) about SE in practice; (2) Lack of connection with industry; and (3) Wrong identification of research problems. The top three suggestions for improving research relevance are: (1) Using appropriate research approaches such as action-research; (2) Choosing relevant (practical) research problems; and (3) Collaborating with industry. By synthesizing all the discussions on this important topic so far, this paper aims to encourage further discussions and actions in the community to increase our collective efforts to improve the research relevance. Furthermore, we raise the need for empirically-grounded and rigorous studies on the relevance problem in SE research, as carried out in other fields such as management science.

## TABLE OF CONTENTS

# 1 INTRODUCTION

Concerns about the state of the practical relevance of research are shared across all areas of science, and Software Engineering (SE) is no exception. A paper in the field of management sciences [55] reported that: "*How and to what extent practitioners use the scientific results of management studies is of great concern to scholars and has given rise to a considerable body of literature*". The topic of research relevance in science in general, also referred to as the "relevance problem", is at least 50 years old, as there are papers dating back to the 1960's, e.g., a paper having the following title: "*The social sciences and management practices: Why have the social sciences contributed so little to the practice of management?*" [56].

David Parnas was one of the first to publish experience-based opinions about questionable relevance of SE research as early as in the 1980's. In his 1985 paper [57], Parnas argued that: "*Very little of [SE research] leads to results that are useful. Many useful results go unnoticed because the good work is buried in the rest*". In a 1993 IEEE Software paper [30], Colin Potts wrote that: "*as we celebrate 25 years of SE, it is healthy to ask why most of the research done so far is failing to influence industrial practice and the quality of the resulting software*".

In around the 25th-anniversary of SE, Robert Glass published an IEEE Software paper [41] in 1994, entitled "*The software-research crisis*". The paper argued that most research activities at the time were not (directly) relevant to practice. Glass also posed the following question: "*What happens in computing and software research and practice in the year 2020?*". Furthermore, he hoped in the paper, for year 2020: "*Both researchers and practitioners, working together, can see a future in which the wisdom of each group is understood and appreciated by the other*" [41].

In a 1998 paper [33], David Parnas said: "*I have some concerns about the direction being taken by many researchers in the software community and would like to offer them my (possibly unwelcome) advice*", and one of those pieces of advice being: "*Keep aware of what is actually happening [in the industry] by reading industrial programs [resources]*". He also made the point that: "*most software developers [in industry] ignore the bulk of our research*".

More recently, some critics have said that: "*SE research suffers from irrelevance. Research outputs have little relevance to software practice*" [21], and that "*practitioners rarely look to academic literature for new and better ways to develop software*" [18]. Also SE education has also been criticized, e.g., "*… SE research is divorced from real-world problems (an impression that is reinforced by how irrelevant most popular SE textbooks seem to the undergraduates who are forced to wade through them)*" [58]. Another team of researchers and practitioners wrote a joint blog post about SE research relevance in which they argued that [13]: "*Some [practitioners] think our field [SE] is dated, and biased toward large organizations and huge projects*".

We have now celebrated 50 years of SE [59], climaxing with the ICSE 2018 conference in Gothenburg, Sweden[1]. Thus, it is a good time to reflect back and wonder to what extent old critique of SE research relevance is still valid. Furthermore, we are close to 2020, i.e., the year targeted by Glass' vision of when "*Software practice and research [would] work together*" [41] (for higher industrial relevance of SE research) have been realized. But has this really happened in a large scale?

Glass had hoped that things would change (improve): "*the gradual accumulation of enough researchers expressing the same view [i.e., the software-research crisis,] began to swing the field toward less arrogant and narrow, more realistic approaches*" [41]. However, as this MLR study would reveal, it could be argued that perhaps we as a community have had only a bit of improvement in terms of research relevance. According to the panelists of an industry-academic panel in ICSE 2011, there is a "*near-complete disconnect between software research and practice*" [46].

---

[1] https://www.icse2018.org/info/50-years-of-se

The argument about practical (ir)relevance of research is not specific to SE, and the issue has been discussed widely in other circles of science, e.g., [60-65]. An online article [60] reported that most works of academics are not shaping industry and the public sector. But instead, "*their [academics] work is largely sitting in academic journals that are read almost exclusively by their peers*". Some of the paper titles on this subject, in other fields of science, look interesting and even bold, e.g., "*Rigor at the expense of relevance equals rigidity*" [61], "*Which should come first: Rigor or relevance?*" [62], "*Reconciling the rigor-relevance dilemma*" [63], "*Information systems research that really matters: Beyond the IS rigor versus relevance debate*" [64], and "*Having it all: Rigor versus relevance in supply chain management research*" [65].

In summary, since as early as 1985, many wake-up calls have been published in the community to reflect on the relevance of SE research [20]. While there have been some positive changes on the issue in recent years, e.g., establishing "industrial" tracks in several SE conferences, many believe we are still far from the ideal situation with respect to research relevance.

The goal of this paper is to synthesize the discussions and arguments in the SE community on the industrial relevance of SE research. To achieve that goal, we report a Multi-vocal Literature Review (MLR) on a set of 54 sources, 36 of which being papers from the peer-reviewed literature and 18 sources from the grey literature (GL), e.g., blog posts and white papers. An MLR [66] is an extended form of a Systematic Literature Review (SLR) which includes the GL in addition to the published (formal) literature (e.g., journal and conference papers). MLRs have recently increased in popularity in SE, as many such studies have recently been published [66-73].

The contributions of this study are novel and useful since, instead of relying on and considering the experience-based opinions mentioned in each of the papers in this area, our MLR collects and synthesizes the opinions from all the sources on this issue and thus it provides a more "holistic" view on the subject. Similar SLRs have been published, each synthesizing the issue in a certain research discipline outside SE, e.g., [74] reported a SLR of the issue in management and human-resource development, [75] reported a systematic analysis of the issue in management science, and [76] reported a review of the issue in the Information Systems (IS) community.

It is worthwhile to clarify the focus of this MLR paper. We focus in this work on research "relevance", and not research "impact", nor "technology transfer" of SE research. As we discuss in Section 2.3, while these concepts (terms) are closely related, they are not the same. Research relevance does not necessarily mean (result in) research impact or technology transfer. A paper / research idea, with high research relevance, has a higher potential for usage (utility) in industry, and could lead to a higher potential (chance) of research impact. The last two concepts (research utility and research impact) are parts of technology transfer phases [77].

The remainder of this paper is structured as follows. Background and a review of the related work are presented in Section 2. We present the design and setup of the MLR in Section 3. The results of the study are presented by answering two review questions (RQs) in Sections 4 and 5. In Section 6, we summarize the findings and discuss the recommendations. Finally, we conclude the paper in Section 7 and discuss directions for future work.

Before going in-depth in the paper, given the controversial nature of the topic of research relevance, the authors should clarify that they are not biased towards finding evidence that some or most of SE research is irrelevant. Instead, our objective and goal in this paper is to synthesize the SE community's discussions on this topic in the last many years. The synthesized evidence may happen or may not happen to support the hypothesis that some or most of SE research is relevant or irrelevant. Such an objective is not the focus of our study.

## 2 BACKGROUND AND RELATED WORK

To set the stage for the rest of the paper, we first use the literature to provide definitions of the following important terms: research relevance, utility, applicability, impact, and rigor, and we then characterize the relationships among them. Then, in terms of related work, we review the literature on research relevance in other disciplines. Finally, we provide a review on the current state of affairs between SE practice and research (industry versus academia), the understanding of which is important to discuss research relevance in the rest of the paper.

> **Note:** To ensure a solid foundation for our paper, we provide a comprehensive background and related work section. If the reader thinks s/he is familiar with precise definitions of research relevance and its related important concepts (research utility, applicability, impact, and rigor), s/he can bypass Section 2 and go directly to Section 3 or 4, to read the design aspects or results of our MLR study.

## 2.1 Understanding the concepts related to research relevance: research utility, applicability, impact, and rigor

To ensure preciseness of our discussions in this paper, it is important to clearly define the terminologies used in the context of this work, which include research "relevance" and its related topics, e.g., research "utility", "impact" and "rigor". We review the definitions of these terms in the literature first. We finally synthesize a definition for research relevance in SE.

### 2.1.1 Research utility and usefulness

Research relevance closely relates to "utility" of research. According to the Merriam-Webster Dictionary, "utility" is defined as "*fitness for some purpose or worth to some end*" and "*something useful or designed for use*".

When research is useful and could provide utility to practitioners, it is generally considered relevant [78]. There have been studies assessing utility of research, e.g., [78] proposed several propositions for utility evaluation of academic research, e.g., (1) utility is dependent not only on academic research supply of knowledge and technology, but equally importantly on demand from industry; and (2) the framework for evaluating research utility must take into consideration differences between research areas.

A 1975 paper entitled "*Notes on improving research utility*" [79] in medical sciences, argued that: "*Researchers are fond of saying that all information is relevant and that more information leads to better outcomes*". But if practitioners' resources (such as time) are limited, practitioners are really interested in the information that provide the most utility to them. When practitioners find most research papers of low utility, they stop paying attention to research papers in general.

### 2.1.2 Research impact

Andreas Zeller, an active SE researcher, defined "impact" as: "*How do your actions [research] change the world?*" [19]. In more general terms, research impact often has two aspects: academic impact and industrial impact. Academic impact is the impact of a given paper on other future papers and activities of other researchers. It is often measured by citations and is studied in bibliometric studies, e.g., [80, 81]. The higher the number of citations to a given paper, the higher its academic impact.

Industrial impact is, however, harder to measure since it is not easy to clearly determine how many times and to what extent a given paper has been read and its ideas have been adopted by practitioners. The "Impact" project [82], launched by ACM SIGSOFT, aimed to demonstrate the (indirect) impact of SE research on industrial SE practices through a number of articles by research leaders, e.g., [83, 84]. Industrial impact of a research paper (irrespective of how difficult it is to assess) could be an indicator of its utility and relevance.

### 2.1.3 Research rigor

Rigor in research refers to "*the precision or exactness of the research method used*" [85]. Rigor can also mean: "*the correct use of any method for its intended purpose*" [86]. In the literature, relevance is often discussed together with research "rigor", e.g., [2, 61, 62]. For example, there is a paper with this title: "*Reconciling the rigor-relevance dilemma*" [63]. A few researchers have mentioned bold statements in this context, e.g., "*Until relevance is established, rigor is irrelevant. When relevance is clear, rigor enhances it*" [87], denoting that there is little value in highly-rigorous, but less-relevant research.

In "*Making research more relevant while not diminishing its rigor*", Robert Glass mentioned that: "*Many believe that the two goals [rigor and relevance] are almost mutually incompatible. For example, rigor tends to demand small, tightly controlled studies, whereas relevance tends to demand larger, more realistic studies*" [17].

To assess different combinations of research rigor and relevance, a paper in psychology [88] presented a matrix model, as shown in Table 1. Where methodological rigor is high, but practical relevance is low, the so-called "pedantic" science is generated. It is the belief of the authors and many other SE researchers (e.g., see the pool of papers in Section 3.7) that most SE papers fall in this category. These are studies that are rigorous in their design and analytical sophistication, but yet fail to address the important issue of relevance. Such research usually derives its questions from theory or from existing published studies, "*the sole criterion of its worth being the evaluation of a small minority of other researchers who specialize in a narrow field of inquiry*" [88].

The quadrant where both practical relevance and methodological rigor are high, is termed as pragmatic science. Such work simultaneously addresses questions of applied relevance and does so in a methodologically robust manner. Clearly, we believe that this particular form of research is the form that should dominate our discipline, an opinion which is also stated in other fields, e.g., psychology [88].

Research representing popular science are highly relevant but lacks methodological rigor. [88] elaborated that popular science "*is typically executed where fast-emerging business trends or management initiatives have spawned ill-conceived or ill-conducted studies, rushed to publication in order to provide a degree of legitimacy and marketing support*". Papers in trade magazines in SE and computing often fall in this category. We also believe that most non-peer-reviewed grey literature (GL) materials, such as blog posts and white papers written by SE practitioners, are often under popular science. Given the popularity of online materials and sources among SE practitioners, we can observe that they often find such materials useful for their problems and information needs.

**Table 1- Research rigor/relevance matrix (source: [88])**

| | Low ← Methodological rigor → High | |
|---|---|---|
| **High** Practical relevance | Popular (popularist) science | Pragmatic science |
| **Low** | Puerile science | Pedantic science |

## 2.2 Research relevance

### 2.2.1 Two aspects of research relevance: Academic (scientific) relevance and industrial (practical) relevance

According to the Merriam-Webster Dictionary, something is relevant if it has "*significant and demonstrable bearing on the matter at hand*".

Similar to research "impact" (as discussed above), research relevance has two aspects in general: academic (scientific) relevance [89] and practical (industrial) relevance. A paper in the field of Management Accounting focused on this very distinction [90]. Academic relevance of a research paper or a research project is the degree of its relevance and the value provided by it for the specific academic field [90]. For example, SE papers which propose interesting insights, frameworks, or formalize certain SE topics which are useful to other researchers in future studies are academically relevant and will have academic impact, even if the subject of those studies are not directly related to SE practice. For example, meta-papers such as the work reported in the current paper (an MLR study), studies in the scope of SE "education research", and various guideline papers in empirical SE such as [91-93] are examples of research undertakings which are academically-relevant, but are not expected to have industrial (practical) relevance.

On the other hand, practical relevance of a research paper or a research project is the degree of its relevance and potential value for SE organizations and industry. Since this paper focuses on practical relevance, in the rest of this paper, when we mention "relevance", we refer to industrial (practical) relevance. A good definition for practical relevance was offered in a management science paper [55]: "*Broadly speaking, research results can be said to be practically relevant if they influence management practice; that is, if they lead to the change, modification, or confirmation of how managers think, talk, or act*".

Similar to the matrix model of Table 1 [88], which showed the combinations of research rigor and relevance, we design a simple matrix model illustrating academic relevance versus practical relevance, as shown in Table 2.

**Table 2- Categorization of research relevance: Academic (scientific) relevance versus practical (industrial) relevance**

| | | |
|---|---|---|
| **High**<br><br>**Practical**<br>**relevance**<br><br>**Low** | Q2: Works in practice (theory not considered) **(majority of technical reports in industry)** | Q4: Highly valuable. Best of both worlds (aspects) |
| | Q3: Not valuable | Q1: Works in theory (practical factors not often considered) **(majority of academic papers)** |
| | **Low**      **Academic**      **High**<br>**relevance** | |

In quarter Q1, we find the majority of academic papers in which the ideas are rigorously developed to work in lab settings and practical considerations (such as scalability and cost-effectiveness of a software testing approach) are not considered. If an industrial practitioner or company gets interested in the approaches or ideas presented in such papers, it would be very hard or impossible to apply the proposed approaches. For example, the study in [6] argued that many model-based testing papers in the literature incur more costs than benefits, when one attempts to apply them, and it went further to state that: "*it is important to always state where the models [to be used for model-based testing] come from: are they artificial or did they already exist before the experiments?*".

Quarter Q2 in Table 2 is the case of the majority of technical reports in industry, which report and reflect on approaches that work in practice, but are often shallow in terms of theory behind the SE approaches. In quarter Q3, we have papers and research projects which are low on both academic and practical relevance, and thus have little value from either aspect.

In quarter Q4, we find papers and research projects which are high on both academic and practical relevance, and are conducted using rigorous (and often empirical) SE approaches. We consider such undertakings as highly valuable which should be the ideal goal for most SE research activities. Examples of such research programs and papers are empirical studies of TDD in practice, e.g., [94, 95], and many other papers which are published in top-quality SE venues, such as ICSE, IEEE TSE and the Springer's Empirical Software Engineering journal, e.g., [96]. Many of the top-quality papers published by researchers working in corporate research centers also fall in this category, e.g., a large number of papers from Microsoft Research, e.g., [97, 98], and successful applications of search-based SE in Facebook Research [99]. Other examples are patents filed on practical SE topics, e.g., a patent on combined code searching and automatic code navigation filed by researchers working in ABB Research [100].

### 2.2.2 Value for both aspects of research relevance: academic and practical relevance

There is obviously value for both aspects of research relevance: academic and practical relevance. Papers that have high academic relevance are valuable since the materials presented in them (e.g., approaches, insights and frameworks) will benefit other researchers in future studies, e.g., various guideline papers in empirical SE such as [91-93] have been cited many times and have helped many SE researchers to better design and conduct empirical SE research.

Papers that have high practical relevance can also be valuable since the materials presented in them have high potentials to be useful for and be applied by practitioners. Furthermore, such papers can help other SE researchers better understand industrial practice – thus stimulating additional relevant papers in the future.

### 2.2.3 Dimensions of practical relevance of research

A paper in the Information Systems (IS) domain [101] stated that "*research in applied fields has to be responsive to the needs of business and industry to make it useful and practicable for them*". The study presented four dimensions of relevance in research which deal with the content and style of research papers, as shown in Table 3 [101]. As we can see, the research topic (problem) is one of the most important aspects. The research topic undertaken by a researcher should address real challenges in industry, especially in an applied field such as SE. It should also be applicable (implementable) and consider current technologies. The study [101] also include writing style as a dimension of relevance. But, it is the opinion of the authors that, while SE papers should be written in a way that is easily readable and understandable by professionals, we would not include writing style as a core dimension of relevance. On the other hand, we believe that the first three dimensions in Table 3 are important and will include them in our definition of relevance at the end of this sub-section.

Table 3- Dimensions of practical relevance of research (from [101])

| Category | Dimensions of relevance | Description |
|---|---|---|
| Topic (problem) | Addressing real challenges (problems) | Does the research address the real problems or challenges that are of concern to professionals? |
| | Applicable (implementable) | Does the research produce the knowledge and offer solutions that can be utilized by practitioners? |
| | Current | Does the research focus on the current technologies and business issues? |
| Writing style | Accessible | Is the research paper easily readable and understandable (in terms of tone, style, structure, and semantics) by professionals? |

In a paper in management science [102], Toffel writes: "*I define relevant research papers as those whose research questions address problems found (or potentially found) in practice and whose hypotheses connect independent variables within the control of practitioners to outcomes they care about using logic they view as feasible*". Toffel further mentioned that: "*To me, research relevance is reflected in an article's research question, hypotheses, and implications*". For a researcher to embark on a research project (or a paper) that want to be relevant to practitioners, Toffel suggested proceeding with a project (paper) only if the researcher can answer "yes" to all three of the following questions:: (1) Is the research question novel to academics (academic novelty/relevance)?, (2) Is the research question relevant to practice?, and (3) Can the research question be answered rigorously? Finally, Toffel believes that relevant research should articulate implications that encourage practitioners to act based on the findings. Researchers should therefore state clearly how their results should influence practitioners' decisions, using specific examples when possible and describing the context under which the findings are likely to apply.

"Applicability" is another term that might be considered related to or even a synonym of "relevance". Merriam-Webster defines applicability as "*the state of being pertinent*", with "relevance" listed as a synonym. "Pertinent" is defined as "*having to do with the matter at hand*", which is essentially the same as the definition of "relevance". While "relevance" and "applicability" are synonyms in terms of linguistics (terminology meanings), applicability is one dimension of research relevance [101], as shown in Table 3. A paper could address real challenges, but it may miss considering realistic assumptions in terms of SE approaches [7], or applying it may incur more costs than benefits ("*cure worse than the disease*"), e.g., [6] argued that many model-based testing papers do not consider this important issue by stating that: "*it is important to always state where the models come from: are they artificial or did they already exist before the experiments?*".

The work by Ivarsson and Gorschek [2] is perhaps the only analytical work in SE which proposed a scoring rubric for evaluating relevance. For the industrial relevance of a study, they argued that the realism of the environment in which the results are obtained influences the relevance of the evaluation. Four aspects of evaluations were considered in evaluating the realism of evaluations: (1) Subjects (used in the empirical study); (2) Context (industrial or lab setting); (3) Scale (industrial scale, or toy example); and (4) Research method. A given SE study (paper) would be scored either contributing to relevance (score=1), or not contributing to relevance (score=0) w.r.t. each of those factors; and the sum of the values would be calculated using the rubric. However, in the opinion of the current paper authors, a major limitation (weakness) of that rubric [2] is that it does not include: addressing real challenges and applicability which are two important dimensions of relevance.

### 2.2.4 Our definition of practical relevance for SE research

We use the above discussions from the literature to synthesize and propose dimensions of relevance in SE research, as shown in Table 4. As shown in Table 4, we found two particular sources in the literature [101, 102] valuable, as they provided solid foundations for dimensions of relevance, and we have adopted our suggested dimensions based on them: (1) Focusing on real-world problems; (2) Applicable (implementable); and (3) Actionable implications.

Table 4- Dimensions of practical relevance for SE research

| Dimension | Description | Adopted from |
|---|---|---|

| Focusing on real-world SE problems | Research topic and research question(s) should improve SE practices and/or address problems found (or potentially-found / seen) in practice (presently or in future) | [101, 102] |
|---|---|---|
| Applicable (implementable) | Using this insight: "*Hypotheses should connect independent variables within the control of practitioners to outcomes they care about using approaches they view as feasible*" [102]. Consisting of:<br>• Realistic inputs / constraints<br>• Controllable independent variables<br>• Outcomes practitioners care about<br>• Feasible and cost-effective approaches (w.r.t. cost-benefit analysis) | [101, 102] |
| Actionable implications | The research (study) should provide implications that would encourage practitioners to act based on the findings. | [102] |

We should clarify the phrase "(presently or in future)" in Table 4, as follows. If the research topic or research question(s) of a study aims at improving SE practices and/or addressing problems, "presently" found or seen in practice, we would consider that study to have practical relevance w.r.t. "present" industrial practices. However, it is widely accepted that, just like any research field, many SE researchers conduct "discovery" (ground-breaking) type of research activities, i.e., research which is not motivated by the "current" challenges in the software industry, but rather presenting ground-breaking new approaches to SE which could or would possibly improve SE practices in future, e.g., research on automated program repair which took many years to be gradually used by industry [103]. In such cases, the research work and the paper would be considered of practical relevance, with the notion of "future" focus, i.e., research with *potential* practical relevance in future. In such cases, research may take several years to reach the actual practice, and then it would become practically relevant. Note that research targeting plausible problems of the future might still benefit from extrapolating needs from current SE practice.

In essence, we can say that there is a spectrum for practical relevance instead of having a binary (0 or 1) view on it, i.e., it is not that a given SE paper/project has zero practical relevance. But instead, we can argue that a given paper/project may have potential practical relevance or even may have the potential to have practical relevance, for the case discovery-type of SE research activities that "may" lead to ground-breaking findings. A classic view on the role of research (including SE research) is to explore directions that industry would not pursue, because these directions would be too speculative, too risky, and quite far away from monetization. This notion of research entails that a large ratio of research activities would have little to no impact in the sense that it will not lead to a breakthrough. However, we argue that by ensuring partial relevance at least, SE researchers would ensure that in case they discover a ground-breaking (novel) SE idea, it would have higher chances of impact on practice and society in general. Our analysis in the next sub-section (Figure 1) will show that: higher potential relevance would have higher chances of impact.

Also related to the above category of research emphasis is the case of SE research projects which start by full or partial industrial funding, whose research objective is to explore new approaches to "do things", e.g., developing novel / ground-breaking software testing approaches for AI software. When industrial partners are involved in a research project, the research direction and papers coming out of such projects would, in most cases, be of practical relevance. But one can find certain "outlier" cases in which certain papers from such projects have tended to diverge from the project's initial focus and has explored and presented topics/approaches which are of low practical relevance.

Inspired by a quote from [102], as shown in Table 4, we divide the applicability dimension (second item in in Table 4) into four sub-dimensions: (1) considering realistic inputs/constraints for the proposed SE technique; (2) controllable independent variables; (3) outcomes which practitioners care about; and (4) feasible and cost-effective approaches (w.r.t. cost-benefit analysis).

We believe that, in SE, relevant research usually satisfies the three dimensions in Table 4 and takes either of the following two forms: (1) picking a concrete industry problem and solving it; and (2) working on a topic that meets the dimensions of relevance in Table 4 using systems, data or artifacts which are similar to industry grade, e.g., working on open-software systems.

### 2.2.5 Spectrum of research relevance according to types and needs of software industry

It is clear that there are different types (application domains) in the software industry, e.g., from firms / teams developing software for safety-critical control systems such as airplanes to companies developing enterprise applications, such as Customer Relationship Management (CRM) systems, to mobile applications and to games software. Nature and needs of SE activities and practices for each of such domains are obviously quite diverse and one can think of this concept as a "spectrum" of SE needs and practices which require a diverse set of SE approaches, e.g., testing an airplane control software would be very different than testing an entertainment mobile app. Thus, when assessing practical relevance of a given SE research idea or paper, one should consider the type of domains and contextual factors [38, 104] under which the SE research idea would be applied in. For example, a heavy-weight requirements engineering approach may be *too much* for developing an entertainment mobile app, and thus not too relevant (or cost effective), but it may be useful and relevant for developing an airplane control software.

This exact issue has been implicitly or explicitly discussed in most of the studies included in our MLR study (Section 3). For example, our pool of studies included a 1992 paper, with the following interesting title: "*Formal methods: Use and relevance for the development of safety-critical systems*" [10]. The first sentence of its abstract reads: "*We are now starting to see the first applications of formal methods to the development of safety-critical computer-based systems*". In the abstract of the same paper [10], the authors present the critique that: "*Others [other researchers] claim that formal methods are of little or no use - or at least that their utility is severely limited by the cost of applying the techniques*".

Taking a step back and looking at the bigger picture of practical relevance of SE research, by a careful review of the sources included in our pool of studies (Section 3), we observed that while the authors of some of the studies, e.g., [10, 51], have raised the above issue of considering the type of software industry when assessing research relevance, the other authors have looked at the research relevance issue, from a "generic" point of view, i.e., providing their critical opinions and assessment of SE research relevance w.r.t. an "overall picture" of the software industry. Since the current work is our first effort to synthesize the community's voice on this topic, based on publications and grey literature from 1985 and onwards, we have synthesized all the voices on the SE research relevance. It would be worthwhile to conduct future syntheses on the issue of research relevance in the pool of studies w.r.t. the types and domains of software systems.

## 2.3 Relationships among relevance and the above related terms

By reviewing the literature (above) and our own analysis, we characterize the inter-relationships among these related terms using a causal-loop (causation) diagram [105] as shown in Figure 1. Let us note that understanding such inter-relationships is important when discussing research relevance in the rest of this paper.

More relevance of a research endeavor would often increase its applicability, usability, and its chances of usage (utility) in industry, and those in turn will increase its chances for (industrial) usefulness and impact. As shown in Figure 1, the main reward metric of academia has been academic impact (citations) which, as we will see in Section 3, is one main reason why the SE research, and almost all other research disciplines (as reviewed in Section 2.2.1), unfortunately suffer from low relevance.

A researcher may decide to work on industrially-relevant or -important topics and/or on "academically-hot" ("academically-challenging") topics. It should be noted that the above options are not mutually exclusive, i.e., one can indeed choose a research topic which is industrially-relevant and important and academically-challenging.

As per our experience, and also according to the literature (Sections 2.4), often times however, aligning these two (selecting an industrially-relevant and an academically-challenging topic) is unfortunately hard [17, 63, 106]. In a book entitled *Administrative Behavior* [107], Herbert Simon used the metaphor of "pendulum swings" to describe the trade-off between the two aspects and compared the attempt to combine research rigor and relevance to an attempt to mix oil with water.

Thus, it seems that relevance and rigor often negatively impact each other, i.e., the more practically relevant a research effort, it could become less rigorous. However, this is not a hard rule and researchers can indeed conduct research that is both relevant and rigorous. For example, one of the authors collaborated with an industrial partner to develop and deploy a multi-objective regression test selection approach in practice [108]. A search-based approach based on genetic algorithms was "rigorously" developed. The approach was also clearly practically-relevant since it was developed using the action-research approach, based on concrete industrial needs. Furthermore, the approach had high industrial "impact" since it was successfully deployed and utilized in practice.

As shown in Figure 1, our focus in this MLR paper is on research "relevance", and not research "impact", nor "technology transfer" of SE research. As it is depicted in Figure 1, while these concepts (terms) are closely related, they are not the same. Research relevance does not necessarily mean (result in) research impact or technology transfer. A paper / research idea, with high research relevance, has a higher potential for usage (utility) in industry, and that could lead to a higher potential (chance) of research impact. The last two concepts (research utility and research impact) are parts of technology transfer phases [77].

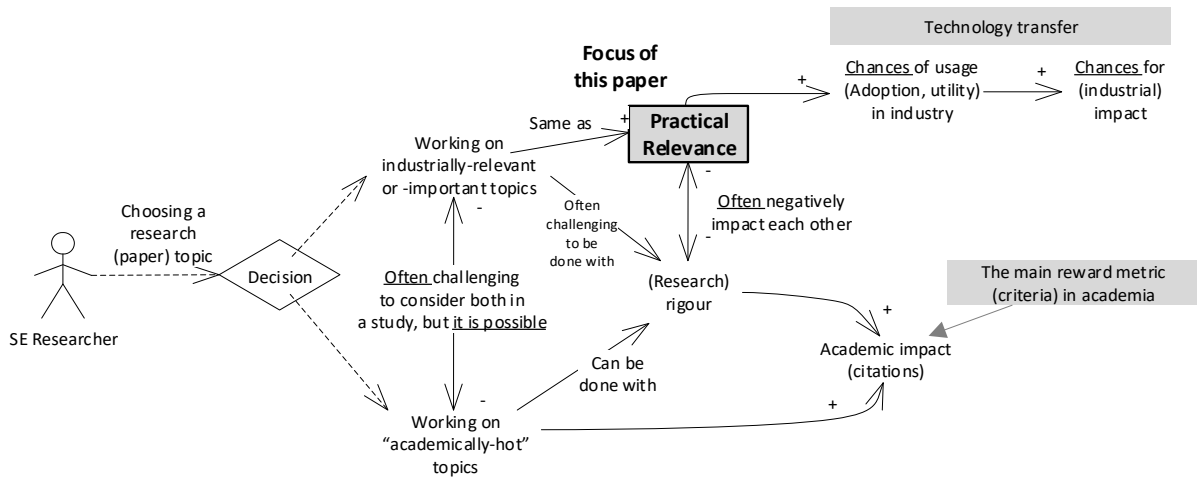

**Figure 1- A causation diagram of relevance and the related terms. Solid arrows indicate causation**

When a researcher chooses to work on an "academically-hot" topic, it is possible to do it with high rigor. This is often because many "simplifications" must be made to formulate (model) the problem from its original industrial form into an academic form [109]. However, when a researcher chooses to work on an industrially-relevant topic, especially when in the form of an Industry-Academia Collaboration (IAC) [106, 110, 111], simplifications cannot be (easily) made to formulate the problem, and thus often times, it becomes challenging to conduct the study with high rigor [109]. When seeing a paper which is both relevant and rigorous, it is important to analyze the "traits" in the study (research project) from inception to dissemination which contribute to its relevance and rigor, and how, as discussed next.

In the opinion of the authors, one main factor which could lead to high research relevance is active IAC [106, 110, 111]. One of the most important aspects in this context is the collaboration mode (style), or degree of closeness between industry and academia, which could have important influence on relevance. One of the best models in this context is the one proposed by Wohlin [112], as shown in Figure 2. There are five levels in this model, which can also be seen as a "maturity" level: (1): not in touch, (2): hearsay, (3): sales pitch, (4): offline, and (5): one team. In levels 1-3, there is really no IAC, since researchers working in those levels only identify a general challenge of the industry, develop a solution (often too simplistic and not considering the context [109]), and then (if operating in level 3) approach the industry to try to convince them to try/adopt the solution. However, since such techniques are often developed with a lot of simplifications and assumptions which are often not valid in industry [109, 113], they often result in low applicability in practice and thus low relevance in industry.

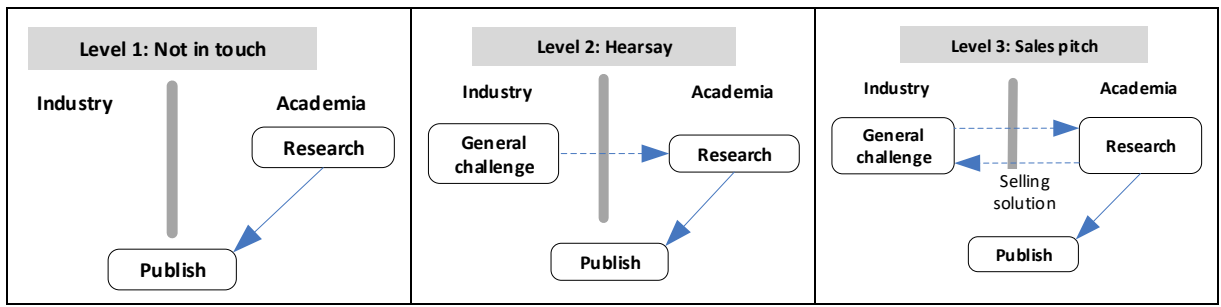

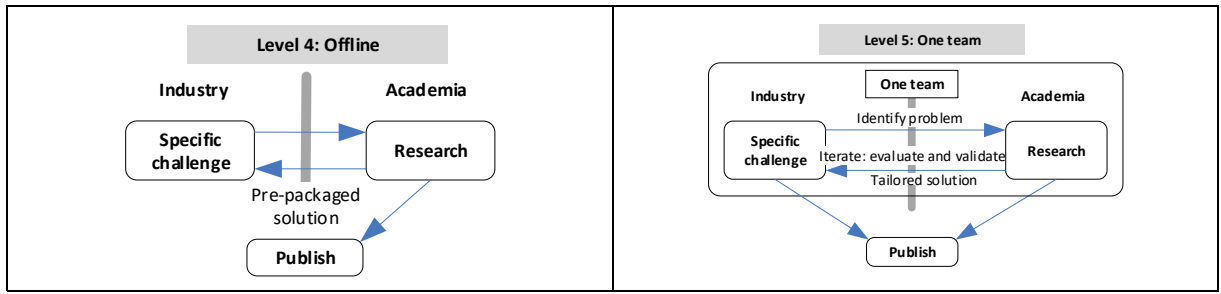

**Figure 2- Five (maturity) levels of closeness between industry and academia (source: [112])**

The Need for Speed (N4S) project, which was a large industry-academia consortium in Finland (n4s.dimecc.com/en) and ran between 2014-2017, is an example of level 5 collaboration. An approach for continuous and collaborative technology transfer in SE was developed in this four-year project. The approach supported "real-time" industry impact, conducted continuously and collaboratively with industry and academia working as one team [114].

Referring back to the "traits" in the study (research project) from inception to dissemination which contribute to its relevance, we have seen first-hand that if a paper comes out of the collaboration mode in level 5 ("one team"), it could have good potential to have high relevance. Of course, when conducting highly relevant research, one should not overlook "rigor", since there is a risk in conducting highly relevant research that lacks methodological rigor [88] (referred to as "popular science"). To ensure research rigor, SE researchers should ensure following the highest levels of rigor using established research methods in empirical SE [91, 115, 116].

## 2.4 Related work

We review next the literature on research relevance in other disciplines, and then the existing systematic review studies on the topic of relevance in SE research.

### 2.4.1 An overview of the "relevance" issue in science in general

To find out about the publications on research relevance in other disciplines, we conducted database searches in the Scopus search engine (www.scopus.com) using the following keywords: "research relevance", "relevance of research", and "rigor and relevance". We observed that the topic of research relevance has indeed been widely discussed in other disciplines, as we were able to get as results a set of several hundred papers. We discuss below a selection of the most relevant papers, identified by our literature review.

The Information Systems (IS) community seems to be particularly engaged in the relevance issue, as we saw many papers in that field. A paper published in the IS community [117] defined a quantitative metric called "relevance coefficient", which was based on the four dimensions of relevance as we discussed in Section 2.1.4 (defined in [101]). The paper then assessed a set of papers in three top IS journals using that metric.

A "relevancy manifesto" for IS research was published [118] in 1999 in the Communications of the AIS (Association for Information Systems), one of the top venues in IS. The author argued that: "*Many practitioners believe academic IS research is not relevant. I argue that our research, and the underlying rewards system that drives it, needs to respond to these concerns*". The author then proposed "*three different scenarios of where the IS field could be 10 years from now [1999]*":

- Scenario 1: Minimal adaptation. The IS field is shrinking, largely due to competition from newly established schools of information technology.
- Scenario 2: Moderate adaptation.
- Scenario 3: High adaptation. The IS field is larger than before, growing in proportion to the demand for graduates with IT skills.

The author said that scenario 1 is the "do nothing" alternative. Scenarios 2 and 3 represent substantial improvements, but they would not occur unless the community acts vigorously to improve the position. We are not sure if any recent study has looked at those three different scenarios to assess which one has been the case in recent years.

A study [119] in management science argued that: "*The relevance literature often moans that the publications of top-ranked academic journals are hardly relevant to managers, while actionable research struggles to get published*". A 1999 paper in the IS community [120] took a philosophical view on the issue and argued that: "*It is not enough for senior IS researchers to call for relevance in IS research. We must also call for an empirically grounded and rigorous understanding of relevance in the first place*".

Another paper in the IS community [121] mentioned that as an applied discipline, is will not achieve legitimacy by the rigor of its methods or by its theoretical base, but by being practically useful. "*Its success will be measured by its contribution to the IS profession, and ultimately to society*" [121]. It further talked about "*the dangers of excessive rigor*" since excessive rigor acts as a barrier to communication with the intended recipients of research results: practitioners, thus it leads to low relevance. The author believed that laboratory experiments are in fact a poor substitute for testing ideas in organizational contexts using real practitioners. "*However, it is much more difficult to do ¡rigorous' research in a practical setting*" [121]. The author also believed that most researchers focus on problems that *can* be researched (using "rigorous" methods) rather than those problems that *should* be researched. In the SE community, Wohlin has referred to this phenomenon as "research under the lamppost" [31].

In other fields, there are even books on relevance of academic conferences, e.g., a book entitled "*Impacts of mega-conferences on the water sector*" [122], which reported that: "*…except for the UN Water Conference, held in Argentina in 1977, the impacts of the subsequent mega-conferences have been at best marginal in terms of knowledge generation and application, poverty alleviation, environmental conservation and /or increasing availability of investments funds for the water sector*".

In the literature of research relevance, we also found a few papers from sub-areas of Computer Sciences, different than SE, e.g., in the field of data-mining [123], Human-Computer Interaction (HCI) [124], and Decision-support systems [125]. Entitled "*Towards more relevance-oriented data mining research*", [123] argued that the data-mining community has achieved "*fine rigor research results*" and that "*the time when DM [data-mining] research has to answer also the practical expectations is fast approaching*".

Published in ACM Interactions magazine, one of the top venues for HCI, [124] started with this phrase: "*Oh, research is research, and practice is practice, and never the twain shall meet*". The paper argued that: "*Some researchers proudly state they are unconcerned with the dirty, messy, unsavory details of commercialization while also complaining that practitioners ignore them. Practitioners deride research results as coming from a pristine ivory tower — intensite perhaps, but irrelevant for anything practical*". Thus, it seems that the issues of research relevance as discussed in SE community does also exist in the HCI community. [124] proposed an interesting idea: "*Between research and practice a third discipline must be inserted, one that can translate between the abstractions of research and the practicalities of practice. We need a discipline of translational development. Medicine, biology, and the health sciences have been the first to recognize the need for this intermediary step through the funding and development of centers for translational science*".

Overall, the extensive discussion on the relevance issue (even referred to as the relevance "problem" [55]) in science in general has been ongoing for more than 50 years, i.e., there are papers published back in the 1960's, e.g., a paper with the following title: "*The social sciences and management practices: Why have the social sciences contributed so little to the practice of management?*" [56]. The papers on the issue have been either often based on opinion, experience, or empirical data (survey studies). Some studies also such as [126] have used related theories such as: Relevance Theory [127], Cognitive Dissonance Theory, and Cognitive Fit Theory for their discussions on this topic. Such studies have aimed at broadening the framing for the issue and provide greater specificity in the discussion of factors affecting relevance.

Many studies have also proposed recommendations (suggestions) for improving research relevance [119, 120, 125], e.g., changes in the academic system (e.g., hiring and tenure committees assigning values for efforts beyond just papers), changes in the research community (e.g., journals appreciating empirical/industrial studies), and changes to the funding system (e.g., funding agencies encouraging further IAC).

Furthermore, it seems that some members of each scientific community seem to be forefront "activists" on the issue while other researchers tend to still put more emphasis on rigor and not consider relevance a major issue [55]. Even, there are reports indicating that the arguments on the issue have sometimes become "heated", e.g., "*… some feelings of unease are to be expected on the part of those scholars whose work is conducted without direct ties to practice*" [128]. Some researchers (e.g., [129]) even think that the "rigor–relevance gap" is "unbridgeable". For example, Kieser and Leiner [129] said: "*From a system theory perspective, social systems are self-referential or autopoietic, which means that communication elements of one system, such as science, cannot be authentically integrated into communication of other systems, such as the system of a business organization*". Such studies sometimes have led to follow-up papers which have criticised such views, e.g., [128].

### 2.4.2 Existing review studies on the issue of research relevance in other fields

In addition to the "primary" studies on this issue, many review studies (e.g., systematic reviews) have also been published in the context of other fields. We did not intend to conduct a systematic review on those review studies, i.e., a tertiary study, instead we show a summary of a few selected review studies. Table 5 lists the selected studies, and we briefly discuss each of them below.

A 1999 systematic mapping paper mapped the research utilization field in nursing [130]. It was motivated by the fact that: "*the past few years [before 1999] have seen a surge of interest in the field of research utilization*". The paper offered critical advice on the issue in that field, e.g., "*Much of it [the literature in this area] is opinion and anecdotal literature, and it has a number of characteristics that suggest the profession has not yet been able to realize sustained initiatives that build and test theory in this area*".

A review paper in management science [55] focused on "*turning the debate on relevance into a rigorous scientific research program*". It argued that "*in order to advance research on the practical relevance of management studies, it is necessary to move away from the partly ideological and often uncritical and unscientific debate on immediate solutions that the programmatic literature puts forward and toward a more rigorous and systematic research program to investigate how the results of scientific research are utilized in management practice*". The paper [55] reviewed 287 papers focusing on the relevance issue in management science and synthesized the reasons for the lack of relevance, and suggested solutions in the literature to improve research relevance. The current MLR is a similar study in the context of SE. The paper [55] organized the literature of the so-called "relevance problem" in management science into several "streams of thought", outlining the causes of the relevance problem that each stream identifies, the solutions it suggests, and summarizes the criticism it has drawn. Those "streams of thought" include: popularization view, institutional view, action-research, Mode-2 research, design science, and evidence-based management. The popularization view is considered the most traditional approach in the programmatic relevance issue. Proponents of this view are concerned with how existing academic knowledge can be transferred to practitioners. They regard the inaccessibility of research and the use of academic jargon as the most important barriers to relevance.

**Table 5- A summary of a few selected review studies on the issue of research relevance in other fields.**

| Study | Fields | Publication year | Years covered | Number of papers reviewed |
|---|---|---|---|---|
| Mapping the research utilization field in nursing [130] | Nursing | 1999 | 1972-1998 | 70 |
| The practical relevance of management research: turning the debate on relevance into a rigorous scientific research program [55] | Management science | 2015 | 1964-2014 | 287 |
| Is management research relevant? A systematic analysis of the rigor-relevance debate in top-tier journals [75] | Management science | 2017 | 1994-2003 | 253 |
| Research–practice gap in applied fields: an integrative literature review [74] | Management, applied psychology, and Human-resource development | 2017 | 1991-2014 | 221 |
| A review of the practical relevance of IS strategy scholarly research [76] | "IS strategy", a sub-field under IS. It is the use of IS to support business strategy. | 2019 | 2008-2018 | 109 |

An SLR of the rigor-relevance debate in top-tier journals in management research was reported in [75]. By being quite similar to [55], it identified four typical positions on rigor and relevance in management research: (1) gatekeepers' orthodoxy, (2) collaboration with practitioners, (3) paradigmatic shift, and (4) refocusing on common good. It argued that, although contradictory, these positions coexist within the debate and are constantly being repeated in the field. The paper linked the findings to the literature on scientific controversies and discussed their implications for the rigor-relevance debate [75].

A literature review of the research–practice gap in the fields of management, applied psychology, and human-resource development was reported in [74]. The paper synthesized the community's discussions on the topic across the above three fields into several themes, e.g., the researching practitioner and the practicing researcher, and engaged scholarship.

A recent 2019 paper [76] reported a descriptive review of the practical relevance of research in "IS strategy", which is a sub-field under IS. This sub-field deals with use of IS to support business strategy [76]. The review paper presented a framework of practical relevance with following dimensions: (1) potential practical relevance, (2) relevance in topic selection, (3) relevance in knowledge creation, (4) relevance in knowledge translation, and (5) relevance in knowledge dissemination.

### 2.4.3 Existing review studies on the topic of research relevance in SE

There has been "meta-literature", i.e., review (secondary) studies, on the research relevance in sub-topics of SE, e.g., [131-134]. The work reported in [131] was an SLR on empirical studies of Agile SE. One of the studied aspects was how useful the reported empirical findings are to the software industry and the research community, i.e., it studied both academic and industrial relevance.

SLR studies such as [132-134] used a previously-reported method for evaluating rigor and industrial relevance in SE [2], as discussed in Section 2.1, which includes the following four aspects for assessing relevance: subjects, context, scale, and research method. As discussed in Section 2.1.4, one major limitation (weakness) of that rubric [2] is that it does not include: addressing real challenges and applicability, which are two important dimensions of relevance, in our opinion.

For example, the SLR reported in [134] reviewed a pool of 58 empirical studies in the area of web application testing, using the industrial relevance and rigor rubrics, as proposed in [2]. The pair-wise comparison of rigor and relevance for the analyzed studies is shown as a bubble-chart in Figure 3. The SLR [134] reported that one can observe quite a reasonable level of rigor and low to medium degree of relevance, in this figure.

By searching the literature, we found no review studies on the topic of research relevance in the "entirety" of the SE research field.

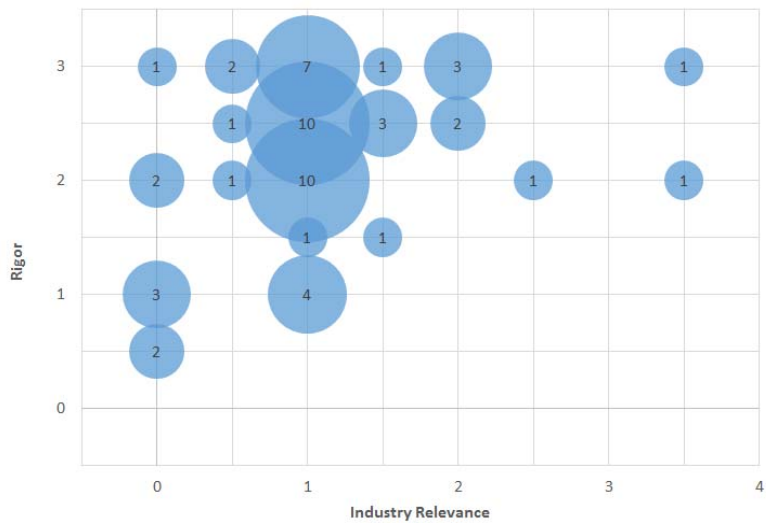

**Figure 3- Rigor versus relevance of the empirical studies as analyzed in the SLR on web application testing [134]**

As another related work is an SLR on approaches, success factors, and barriers for technology transfer in SE [77]. As discussed in Section 2.3, research relevance, and technology transfer of SE research are closely-related concepts. Among the findings of the SLR was that, empirical evidence, maturity, and adaptability of the technology (developed in research settings) seem important preconditions for successful transfer, while social and organizational factors seem important barriers to successful technology transfer. We could contrast the above factors with dimensions of research relevance, in which one important factor is focusing on real-world SE problems (Table 4).

### 2.5 Current state of affairs: SE practice versus research (industry versus academia)

To better understand research relevance, we also need to have a high-level view of the current state of affairs between SE practice versus research (industry versus academia) [106, 111, 135]. It is without a doubt that the current level of IAC in SE is relatively small compared to the level of activities and collaborations within each of the two communities, i.e., industry-to-industry collaborations and academia-to-academia collaborations. A paper [121] in the IS community put this succinctly as follows: "*While they deal with the same subject matter, practitioners and researchers mix in their own circles, with very little communication between them*".

While it is not easy to get quantitative data on IAC, we can look at the estimated population of the two communities. According to a report by Evans Data Corporation [136], there were about 23 million software developers worldwide in 2018, and that number is estimated to reach 27.7 million by 2023. According to a 2012 IEEE Software paper [9], *"4,000 individuals"* are *"actively publishing in major [SE] journals"*, which can be used as the estimated size (lower bound) of the SE research community. If we divide the two numbers, we can see that on average, there is one SE academic for every 5,750 practicing software engineer, denoting that the size of the SE research community is very small compared to the size of the SE workforce. To better put things in perspective, we visualize the two communities and the current state of collaborations in Figure 4.

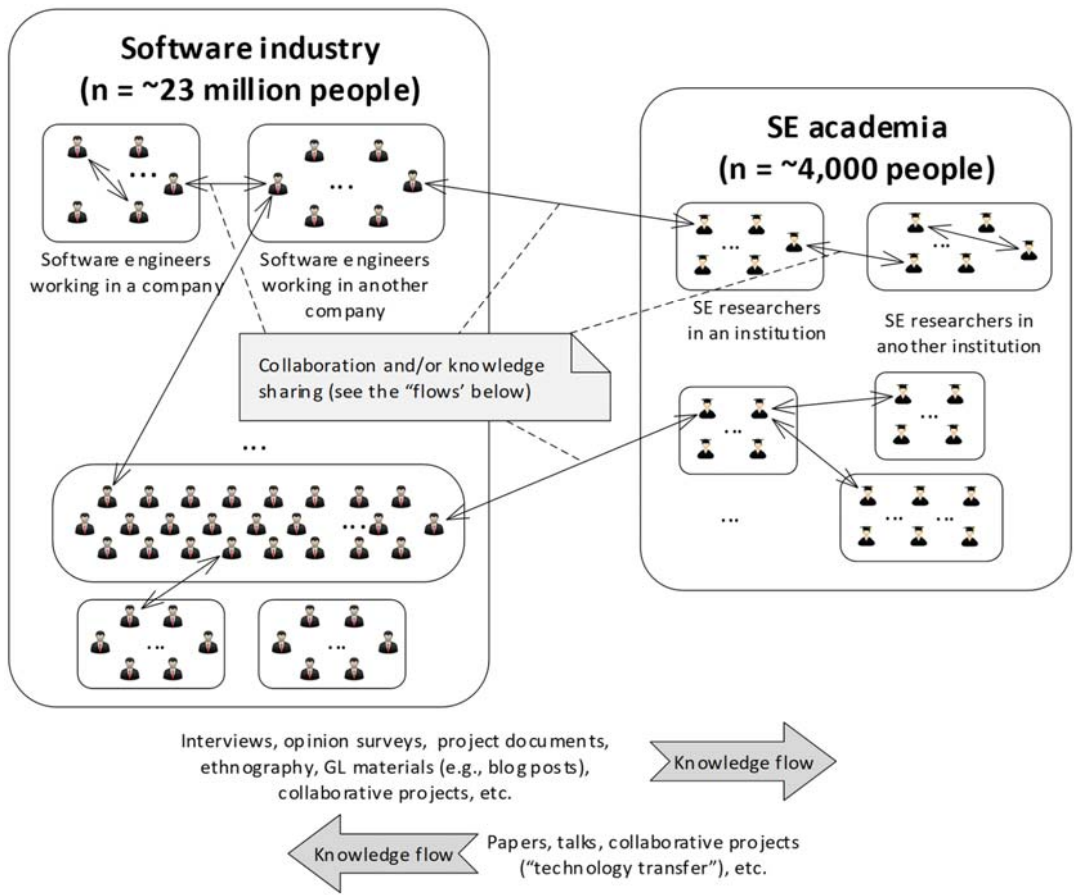

**Figure 4- Visualizing the current state of collaborations in the software industry, academia and between them**

Inside each of the communities in Figure 4, there are groups of software engineers working in each company, and SE researchers in each academic institution (university). The bi-directional arrows (edges) in Figure 4 denote the collaboration and knowledge sharing between members of practitioners and researchers, inside their companies, universities, and with members of the other community. In the bottom of Figure 4, we are showing the knowledge flow in both directions between the two communities. From industry to academia, knowledge flows could occur in various forms, e.g., interviews, opinion surveys, project documents, ethnography, reading GL sources (such as blog posts), and joint collaborative projects. In the other direction, from academia to industry, knowledge flows occur in forms of practitioners reading academic papers, talks by researchers, and joint collaborative projects ("technology transfer"), etc.

While we can see extensive collaborations within each of the two communities (software industry and academia), many believe that the level of collaborations between members of the two communities are much less frequent [121]. There are few quantitative data sources on the issue of interaction and information flow between the two communities. For example, in a survey of software testing practices in Canada [137], a question asked practitioners to rate their frequency of interaction (collaboration) with academics. Based on the data gathered from 246 practitioners, the majority of respondents (56%) mentioned never interacting with the researchers in academia. 32% of the respondents mentioned seldom interactions. Those who interacted with researchers once a year or more only covered a small portion among all respondents (12%). Thus, we see that, in general, there are limited interaction, knowledge exchanges and information flows between the two communities in SE. Nevertheless, we could clarify that there are multiple communities within academic SE and industry SE, respectively. In such a landscape, some communities might collaborate more than others, and some industry sectors are probably closer to research than others.

The weak connection between industry and academia is also visible from "*the difficulties we have to attract industrial participants to our conferences, and the scarcity of papers reporting industrial case studies*" [48]. Two SE researchers who both moved to industry wrote in a blog post [46]: "*While researchers and practitioners may mix and mingle in other specialties, every SE conference seemed to be strongly biased to one side or the other*" and "*…only a handful of grad students and one or two adventurous faculty attend big industrial conferences like the annual Agile get-together*". On a similar topic, there are insightful stories about moving between industry and academia by people who have made the move, e.g., [138].

Of course, things are not all bad, and there have been many positive efforts to bring software industry and academia closer, e.g., many specific events such as panels have been organized on the topic, such as the following example list:

- A panel in ICSE 2000 conference with this title: "*Why don't we get more (self?) respect: the positive impact of SE research upon practice*" [139]
- A panel in ICSE 2011 conference with this title: "*What industry wants from research*" [140]
- A panel in FSE 2016 conference with this title: "*The state of SE research*" [141]
- A panel in ICST 2018 conference with this title: "*When are software testing research contributions, real contributions?*" [53]

While the above panels are good examples of efforts to bring industry and academia closer, they are not examples of successes in terms of *evenly-mixed* attendance by both practitioners and researchers, since most of their attendees were researchers. While some researchers clearly see the need to pursue discussions on this topic, most practitioners in general do not seem to care about this issue. This opinion was nicely summarized in a recent book chapter by Beecham et al. [45]: "*While it is clear that industry pays little attention to software engineering research, it is not clear whether this [relevance of SE research] matters to practitioners.*"

The "Impact" project [82], launched by ACM SIGSOFT, aimed to demonstrate the (indirect) impact of SE research through a number of articles by research leaders, e.g., [83, 84]. Although some impact can certainly be credited to research, we are not aware of any other engineering discipline trying to demonstrate its impact through such an initiative. This, in itself, is a symptom of a lack of impact as the benefits of engineering research should be self-evident.

In a classic book entitled "*Software creativity 2.0*" [142], Robert Glass dedicated two chapters to "*theory versus practice*" and "*industry versus academe*" and presented several examples (which he believes are "*disturbing*") on the mismatch of theory and practice. One section of the book focused especially on "*Rigor vs. relevance [of research]*" (Section 8.8). Another book by Robert Glass was on "*Software conflict 2.0: the art and science of software engineering*" [143] in which he also talked about theory versus practice and how far (and disconnected) they are.

Before leaving this section, we want to clarify that Figure 4 shows a simplified picture as academic research is not limited to the universities. Public and private research institutes as well as corporate research centers publish numerous SE papers. Research institutes focusing on applied science, which may or may not be connected to universities, can help bridging gaps between academia and industry – thus supporting relevant research. The industry connections are often stressed in the corresponding mission statements, e.g., "*partner with companies to transform original ideas into innovations that benefit society and strengthen both the German and the European economy*" (from the mission statement of Fraunhofer family of research institutes in Germany), "*work for sustainable growth in Sweden by strengthening the competitiveness and capacity for renewal of Swedish industry, as well as promoting the innovative development of society as a whole*" (from the mission statement of the Research Institutes of Sweden, RISE), and "*connect people and knowledge to create innovations that boost the competitive strength of industry and the well-being of society in a sustainable way*" (TNO, the Netherlands). Finally, corporate research centers typically must demonstrate practical relevance of the research to motivate their existence. Corporate research centers that frequently publish in SE venues include: Microsoft Research, ABB Corporate Research, and IBM Research.

## 3 RESEARCH METHOD AND SETUP OF THE MLR

We discuss in the following the different aspects of the research method used for conducting the MLR.

### 3.1 Goal and review questions (RQs)

The goal of our MLR is to synthesize the existing literature and debates in the SE community about research relevance. Based on this goal, we raised two review questions (RQs):

- RQ 1: What root causes have been reported in the SE community for the relevance problem (lack of research relevance)?
- RQ 2: What ideas have been suggested for improving research relevance?

### 3.2 Deciding between an SLR and an MLR

During the planning phase of our review study, we had to decide whether to conduct an SLR (by only considering peer-reviewed sources) or to also include the grey literature (GL), e.g., blog posts and white papers, and conduct an MLR [66]. In our initial literature searches, we found several well-argued GL sources about research relevance in SE, e.g., [8, 16, 19, 21], and it was evident that those sources would be valuable for us when answering the study's RQs. We thus determined that we should include GL in our review study, and thus decided to conduct an MLR instead of a SLR.

### 3.3 Overview of the MLR process

MLRs have recently started to appear in SE. According to a literature search [66], the earliest MLR in SE seems to have been published in 2013, on the topic of technical debt [144]. More recently, more MLRs have been published, e.g., on smells in software test code [145], on serious games for software process education [146], and on characterizing DevOps [70].

A recent guideline for conducting MLRs in SE has been published [66], which is based on the SLR guidelines proposed by Kitchenham and Charters [147], and MLR guidelines in other disciplines, e.g., in medicine [148] and education sciences [149]. As noted in [66], certain phases of MLRs are quite different than those of regular SLRs, e.g., searching for and synthesizing grey literature.

To conduct the current MLR, we used the guideline mentioned above [66] and our recent experience in conducting several MLRs, e.g., [67, 71, 145]. We first developed the MLR process as shown in Figure 5. The authors conducted all the steps as a team.

We present the subsequent phases of the process in the following sub-sections: (phase 2) search process and source selection; (phase 3) development of the classification scheme (map); (phase 4) data extraction and systematic mapping; and finally (phase 5) data synthesis. As we can see, this process has a lot of similarity to the typical SLR processes [147] and also SM processes [93, 150], the major difference being only in the handling of the grey literature, i.e., searching for those sources, applying inclusion/exclusion criteria to them and synthesizing them.

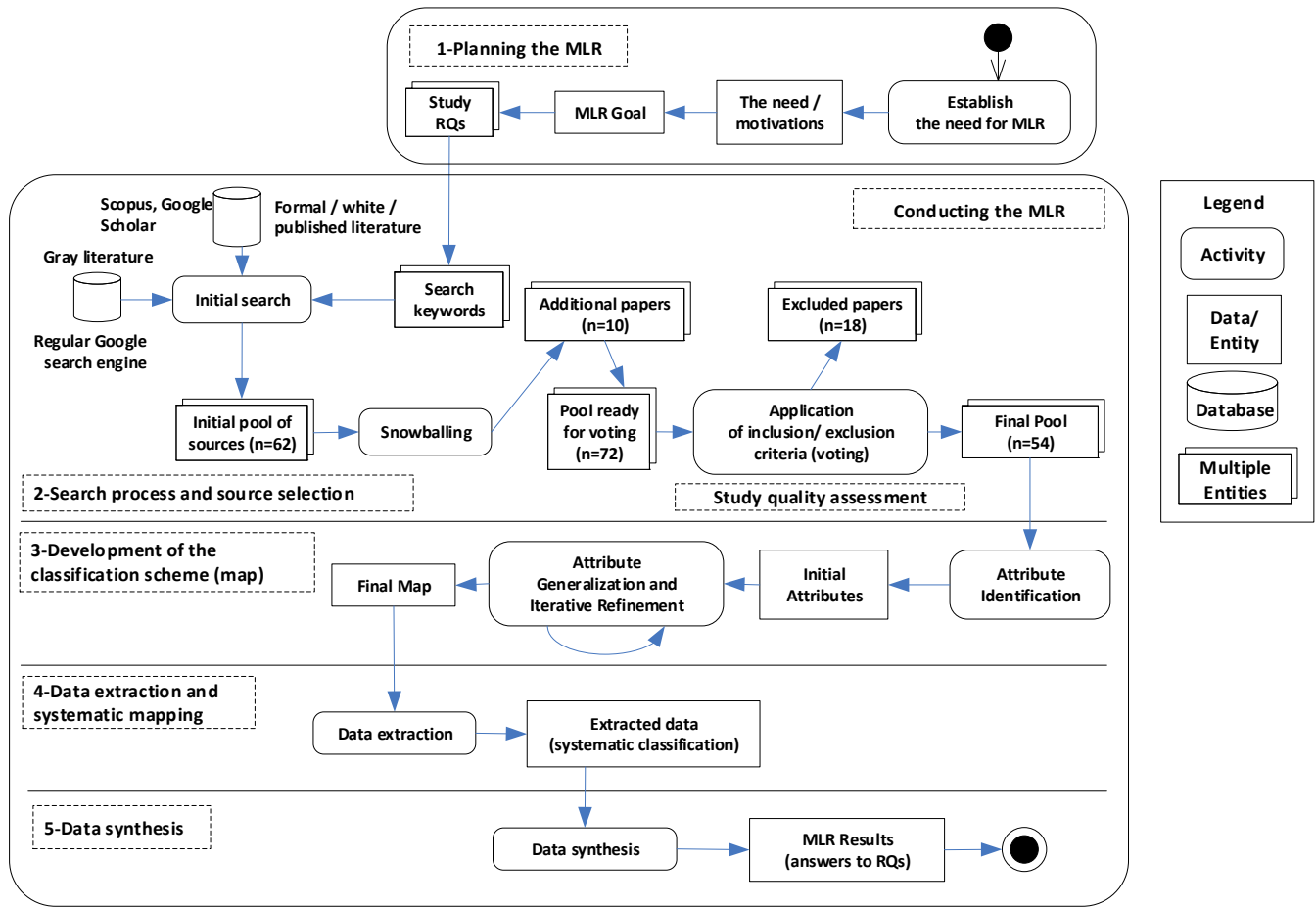

**Figure 5-An overview of our MLR process (as a UML activity diagram)**

## 3.4 Source selection and search keywords

As suggested by the MLR guidelines [66], and also as done in several recent MLRs, e.g., [67, 71, 145], we performed the searches for the formal literature (peer-reviewed papers) using the Google Scholar and the Scopus search engines. To search for the related grey literature, we used the regular Google search engine. Our search strings were: (a) *Relevance software research*; (b) *Relevant software research*; and (c) *utility software research.*

Details of our source selection and search keywords approach were as follows. The authors did independent searches with the search strings, and during this search, they already applied inclusion/exclusion criterion for including only those results which explicitly addressed "relevance" of SE research.

Typically in SM and SLR studies, a team of researchers includes all the search results in the initial pool and then performs the inclusion/exclusion as a separate step. This results in huge volumes of irrelevant papers. For example, in an SLR [151], the team of researchers started with an initial pool of 24,706 articles but out of those only 25 (about 0.1% of the initial pool) were finally found relevant. This means researchers had to spend a lot of (unnecessary) effort due to the very relaxed selection and filtering in the first phase. In line with our recent work [152, 153], we performed rigorous initial filtering to guard against including too many irrelevant papers. On the other hand, we made sure to include both *clearly-related* and *potentially-related* papers in the candidate pool to guard against missing potentially-relevant paper.

To ensure finding the relevant grey literature, we utilized the "relevance ranking" of the Google search engine (i.e., the so-called *PageRank* algorithm) to restrict the search space. For example, if one applies search string (a) above ("*Relevance software research*") to the Google search engine, 293,000,000 results would show as of this writing (February 2019), but as per our observations, relevant results usually only appear in the first few pages. Thus, similar to what was done in several recent

MLRs, e.g., [67, 71, 145], we checked the first 10 pages (i.e., somewhat a search "saturation" effect) and only continued further if needed, e.g., when the results in the 10th page still looked relevant.

As a result of the initial search phase, we ended up with an initial pool of 62 sources. To ensure including all the relevant sources as much as possible, we also conducted forward and backward *snowballing* [154] for the sources in formal literature, as recommended by systematic review guidelines, on the set of papers already in the pool. Snowballing, in this context, refers to using the reference list of a paper (backward snowballing) or the citations to the paper to identify additional papers (forward snowballing) [154]. By snowballing, we found and added to the candidate pool 10 additional sources, bringing the pool size to 72 sources. For example, source [3] was found by backward snowballing of [155], source [5] was found by backward snowballing of [54].

## 3.5 Quality assessment of the candidate sources

An important phase of study selection for an SLR [147] or an MLR study [66] is quality assessment of the candidate sources. Because the topic under study (the relevance of SE research) differs from the topics of "technical" SE papers (e.g., papers in SE topics such as testing), and because we intended to include both peer-reviewed literature and grey literature (GL), we had to develop appropriate quality assessment criteria.

After reviewing the SLR guidelines [147] and MLR guidelines [66], we established the following quality assessment criteria:

- Determine the information each candidate source used as the basis for its argumentation. After reviewing several candidate sources, we found that the argumentations are based on either: (1) Experience or opinions of its authors (opinions were often based on years of experience), or (2) Based on empirical data (e.g., conducting a survey with a pool of programmers about relevance of SE research or a set of SE papers).
- If the source was from the peer-reviewed literature, we would assess the venue the paper has been published in, and the research profile of the authors. We used their citation counts as a proxy of research strength / expertise of the authors.
- If the candidate source was from the grey literature (GL), we used a widely-used [66] quality-assessment checklist, named the AACODS (Authority, Accuracy, Coverage, Objectivity, Date, Significance) checklist [156] for appraising the source.

A given source was ranked higher in quality assessment (to discuss about relevance of SE research), if it was based on empirical data, rather than just experience or opinions. However, when the author(s) of an experience-based or opinion-based source had a high research profile (many citations), then we also ranked it high in quality assessment. Also, we found that for almost all the papers, e.g., [40], authors of experience/opinion papers in the pool had substantial experience working in (or with the) industry (more details in Section 3.8.2). Thus, they had the qualifications / credibility to cast their experience-based opinions in their papers.

Also, venues for published papers provided indications about their quality. In the candidate pool of 62 sources, we found that four papers were published in IEEE Computer, nine papers were published in IEEE Software, two in Communication of ACM, and two in the Empirical Software Engineering journal, all considered to be among the top-quality SE venues.

As examples, we show in Table 6 the log of data that we derived to support quality assessment of eight example candidate sources (seven papers and one GL source). For the GL source example (a blog-post), we determined that all six criteria of the AACODS checklist (authority, accuracy, coverage, objectivity, date, significance) had a value of 1 (true). This determination was justified as follows. The blog-post author is a highly-cited SE researcher (4,084 citations as of Feb. 2019, according to Google Scholar). He had written the post motivated by: "*I just attended FSE 2016, a leading academic conference on software engineering research*". According to his biography (https://dirkriehle.com/about/half-page-bio), this researcher has also been involved in industrial projects and companies in several occasions.

Three of the candidate papers in Table 6 used empirical data as the basis (data source) for their argumentation. For example, [4] used the "participant observation" research method [115] to report a three-year experience of software evolution in a software start-up and its implication for SE research relevance. [7] reported an empirical study with programmers to assess whether automated debugging techniques are actually helping programmers, and then discussed the implication for SE research relevance.

For these eight example sources, we can see that they are all considered to be of high quality, i.e., deserve to be included in the pool of the MLR.

**Table 6- Quality assessment of several candidate sources**

| Paper Title | Source type | | Base (data source) for argumentation | | Quality assessment | | | | | | | |
|---|---|---|---|---|---|---|---|---|---|---|---|---|
| | Peer-reviewed literature (Paper) | Grey literature (GL) | Experience / opinion | Based on empirical data | Comments | Authority | Accuracy | Coverage | Objectivity | Date | Significance |
| A deja-vu look at software engineering researchers who care about practice [1] | 1 | | x | | Published in CACM. Highly cited author: Robert L. Glass. Also based on industrial experience | | | | | | |
| A method for evaluating rigor and industrial relevance of technology evaluations [2] | 1 | | x | | Published in: Empirical Software Engineering journal. Paper cited: 186 | | | | | | |
| A new paradigm for applied requirements engineering research [3] | 1 | | | Gathering opinions of attendees in a workshop at RE'12 | Published in: RE conference | | | | | | |
| A three-year participant observation of software start-up software evolution [4] | 1 | | | Participant observation | Published in: ICSE SEIP. An active SE researcher | | | | | | |
| Albert Einstein and empirical software engineering [5] | 1 | | x | | Published in: IEEE Computer. Author: Shari Lawrence Pfleeger (8,176 citations) | | | | | | |
| An experience report on applying software testing academic results in industry: we need usable automated test generation [6] | 1 | | x | | Published in: Empirical Software Engineering journal | | | | | | |
| Are automated debugging techniques actually helping programmers? [7] | 1 | | | Empirical studies with programmers | Published in: ISSTA. Cited by 325 | | | | | | |
| Challenges to making software engineering research relevant to industry [8] | | 1 | x | | Highly cited SE researcher: 4,084 | 1 | 1 | 1 | 1 | 1 | 1 |

### 3.6 Inclusion / exclusion criteria

The next stage was devising a list of inclusion/exclusion criteria and applying them via a voting phase. We set the inclusion criterion as follows: The source should clearly focus on and comment about SE research relevance, and should possess satisfactory quality in terms of evidence or experience (as discussed in the previous section).

To assess candidate sources for inclusion/exclusion when their titles (main focus) was not "relevance", but one of the related terms (e.g., impact), we used the definition of research relevance from Section 2.1. Any source, which did not clearly focus on "relevance", was excluded. During the inclusion/exclusion process, the authors did the judgements on the candidate sources on a case-by- case basis. For example, we excluded several papers published out of the ACM SIGSOFT "Impact" project [82-84], since they had only focused on research impact, and had not discussed any substantial material about relevance. As another example, a potentially related source was a blog-post in the IEEE Software official blog, entitled "The value of applied research in software engineering" (blog.ieeesoftware.org/2016/09/the-value-of-applied-research-in.html). The post was interesting, but it focused on the value of applied research and impact, and not on "relevance". Including such sources would not have contributed any data towards answering our two RQs (see Section 3.1), since they had not contributed any direct substance about "relevance".

Several papers had the term "relevance" in their title explicitly, e.g., "*Strategies for industrial relevance in software engineering education*" [157], "*The relevance of education to software practitioners*" [158], but were not included since they focused on relevance of SE "education" and not research relevance. Our exclusion criteria were: (1) non-English papers, and (2) full papers being not available.

As shown in Figure 5, application of the inclusion/exclusion criteria and quality assessment resulted in exclusion of 18 sources. To minimize the risk of bias in applying inclusion/exclusion criteria, the first and second authors voted for the papers to be included. The final pool of sources included 54 sources, which we present next, and then review them in the rest of this paper.

### 3.7 Pool of sources

We show the final pool of sources in Table 7 along with their source types. 36 were published literature (papers) and 18 were GL sources. We show the annual trend of sources in Figure 6. We see in Figure 6 that community discussions on the topic have become more active in recent years, after 2010. Implying the increase of discussions on the debate from the increase in the "number" of sources in recent years should be done with "a grain of salt". As we report in the rest of this paper, we should consider the "content" of sources more than the number of sources, but of course the increase in numbers of sources is a message on its own.

We see that the issue was discussed as early as in 1985 in a paper entitled "*On the relevance of formal methods to software development*" [20] (S20). In the rest of this section, we refer to the sources in the form of *Si*, as shown in Table 7.

### Table 7-Review of literature on research relevance in SE

| ID | Paper / source title | Publication year | Reference | Source type | | Data source for argumentation | |
|----|----------------------|------------------|-----------|-------|-----|------------------------|----------------------|
| | | | | Paper | GL | Experience / opinion | Based on empirical data |
| S1 | A Deja-vu look at SE researchers who care about practice | 2007 | [1] | x | | x | |
| S2 | A method for evaluating rigor and industrial relevance of technology evaluations | 2011 | [2] | x | | x | |
| S3 | A new paradigm for applied requirements engineering research | 2013 | [3] | x | | | Gathering opinions of attendees in RE'12 |
| S4 | A three-year participant observation of software start-up software evolution | 2017 | [4] | x | | | Participant observation |
| S5 | Albert Einstein and empirical SE | 1999 | [5] | x | | x | |
| S6 | An experience report on applying software testing academic results in industry: we need usable automated test generation | 2017 | [6] | x | | x | |
| S7 | Are automated debugging techniques actually helping programmers? | 2011 | [7] | x | | | Empirical studies with practitioners |
| S8 | Challenges to making SE research relevant to industry | 2016 | [8] | | x | x | |
| S9 | Embracing the engineering side of SE | 2012 | [9] | x | | x | |
| S10 | Formal methods: use and relevance for the development of safety-critical systems | 1992 | [10] | x | | x | |
| S11 | Grand challenges in program comprehension and software repository mining: interdisciplinary and research relevance | 2018 | [11] | | x | x | |
| S12 | How can requirements engineering research become requirements engineering practice | 1997 | [12] | x | | x | |
| S13 | How do practitioners perceive SE research? | 2011 | [13] | x | | | Interviewing leading practitioners |
| S14 | How practitioners perceive the relevance of ESEM research | 2016 | [14] | x | | | 9,941 ratings from 437 practitioners |
| S15 | How practitioners perceive the relevance of SE research | 2015 | [15] | x | | | 17,913 ratings from 512 practitioners |
| S16 | Impactful SE research: some do's and more don'ts | 2017 | [16] | | x | x | |
| S17 | Making research more relevant while not diminishing its rigor | 2009 | [17] | x | | x | |
| S18 | Making SE research relevant | 2014 | [18] | x | | x | |
| S19 | On impact in SE research | 2018 | [19] | | x | x | |

| S20 | On the relevance of formal methods to software development | 1985 | [20] | x | | x | |
| S21 | On the worthiness of SE research | 2017 | [21] | | x | x | |
| S22 | Practicing what we preach | 2014 | [22] | x | | x | |
| S23 | Really rethinking formal methods | 2010 | [23] | x | | x | |
| S24 | Requirements engineering and technology transfer-obstacles, incentives and improvement agenda | 2002 | [24] | x | | | Summarises the results of two panel discussions |
| S25 | Requirements researchers: do we practice what we preach? | 2002 | [25] | x | | x | |
| S26 | Research and practice: Essays of a researcher-practitioner | 2018 | [26] | | x | x | |
| S27 | Research methodology on pursuing impact-driven research | 2018 | [27] | | | x | |
| S28 | Software engineering-missing in action: A personal perspective | 2010 | [28] | x | | x | |
| S29 | Software engineering research and industry: a symbiotic relationship to foster impact | 2018 | [29] | x | | x | |
| S30 | Software-engineering research revisited | 1993 | [30] | x | | x | |
| S31 | Software engineering research under the lamppost | 2013 | [31] | | x | x | |
| S32 | Software testing and industry needs | 2006 | [32] | x | | x | |
| S33 | Successful SE research | 1998 | [33] | x | | x | |
| S34 | Summary of the Dagstuhl workshop on future directions in SE | 1993 | [34] | x | | | Summary of the Dagstuhl workshop on future directions in SE |
| S35 | The (ir)relevance of academic SE research | 2014 | [35] | | x | x | |
| S36 | The birth of refactoring: a retrospective on the nature of high-impact SE research | 2015 | [36] | x | | x | |
| S37 | The black hole of SE research | 2015 | [37] | | x | x | |
| S38 | The case for context-driven SE research | 2017 | [38] | x | | x | |
| S39 | The other impediment to SE research | 2010 | [39] | x | | x | |
| S40 | The relationship between theory and practice in SE | 1994 | [40] | x | | x | |
| S41 | The software-research crisis | 1994 | [41] | x | | x | |
| S42 | The two paradigms of software development research | 2018 | [42] | x | | x | |
| S43 | The world view of research in software engineering | 2018 | [43] | | x | x | |
| S44 | Towards improved adoption: effectiveness of research tools in the real world | 2016 | [44] | x | | | Experiments to assess effectiveness of research tools in real-world contexts |
| S45 | Triangulating research dissemination methods: a three-pronged approach to closing the research–practice divide | 2018 | [45] | x | | | Surveying practitioners |
| S46 | Two solitudes illustrated | 2012 | [46] | | x | x | |
| S47 | Unappreciated bubble research | 2017 | [47] | | x | x | |
| S48 | Useful SE research: leading a double-agent life | 2011 | [48] | x | | x | |
| S49 | Waiting for the funerals: culture in software engineering research | 2018 | [49] | | x | x | |
| S50 | What do software engineers care about? gaps between research and practice | 2017 | [50] | x | | | Opinion survey of 67 software engineers |
| S51 | What industry needs from architectural languages: A survey | 2013 | [51] | x | | | Interviewing 48 practitioners |
| S52 | What industry wants from academia in software testing? Hearing practitioners' opinions | 2017 | [52] | x | | | Opinion survey of 63 software engineers |
| S53 | When are software testing research contributions, real contributions? | 2018 | [53] | | x | x | |
| S54 | Who are we doing global SE research for? | 2013 | [54] | x | | | Semi-structured interviews with senior managers from ten companies |

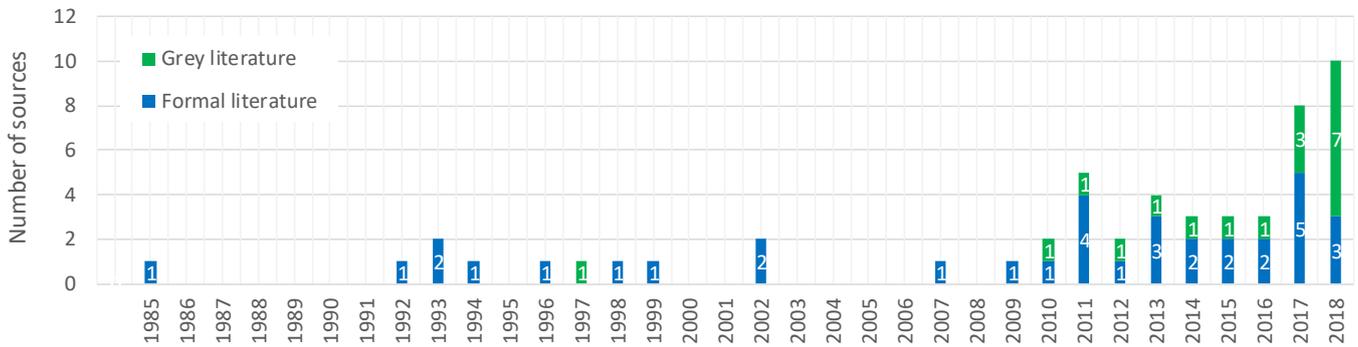

**Figure 6- Annual trend of sources in our MLR pool**

## 3.8 Properties of the sources

Before presenting the results in the light of the RQs, we present two aspects in the pool of studies; (1) SE areas (topics) covered in the sources; and (2) Bases used for argumentation in the sources (experience or empirical data).

### 3.8.1 SE areas (topics) covered in the sources

We classified the 54 sources based on the SE areas (topics) covered in them. While 36 sources were generic (considered SE as a whole), 18 sources discussed research relevance in the context of specific SE areas, e.g., five sources focused on requirements engineering (RE), and four papers focused on testing, three focused on formal methods. For example, S3 presented a new paradigm for "applied" requirements engineering (RE) research, in which relevance was a core. S6 was an experience report on applying software testing academic results in industry, in which the relevance of software testing papers was questioned.

S54 is a paper titled "*Who are we doing global software engineering research for?*". Based on semi-structured interviews with senior managers and project managers from ten companies, it mentioned insightful comments about the relevance problem.

In summary, we can see that most sources have focused on SE as a whole when discussing the relevance problem. The sources focusing on sub-areas of SE are by authors who work in those specific areas and thus have shared their views on the relevance problem with a focus on those specific areas.

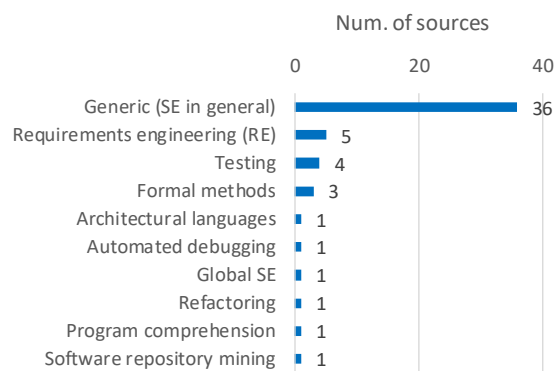

**Figure 7- SE areas (topics) in which research relevance was discussed**

### 3.8.2 Bases used for argumentation in the sources

As discussed in Section 3.5, we wanted to ensure including sources which had used convincing bases for their argumentation. Via a careful quality assessment process on the candidate sources, we verified that aspect (see the examples in Table 6). Among the final pool of sources, 36 sources were based on experience of their authors, and 18 sources were based on empirical data. Details can be found in in Table 7. We discuss a few examples next.

S15 assessed "how practitioners perceive the relevance of SE research". In a detailed questionnaire, the study solicited 17,913 ratings for a large set of SE paper from 512 practitioners in a large company (Microsoft). Similarly, S14 considered the papers published in a SE conference (International Symposium on Empirical Software Engineering and Measurement, ESEM) and solicited 9,941 ratings from 437 practitioners.

Two examples of sources which used experience of their authors for argumentation are S11 and S47. S11 was a keynote talk in the International Conference on Mining Software Repositories (MSR) 2018, and had the title of: *"Grand challenges in program comprehension and software repository mining: interdisciplinarity and research relevance"*. It was given by a highly-cited author (5,716 citations as of Feb. 2019, according to Google Scholar). We found the talk slides insightful, thus we determined all six criteria of the AACODS checklist (authority, accuracy, coverage, objectivity, date, significance) to be one (true) for S11.

S47 is a blog-post entitled *"Unappreciated bubble research"* written by a practitioner who is the author of a book on the C language: (www.knosof.co.uk/cbook/cbook.html). The author has been collaborating / discussing the issue of relevance with researchers for many years (as per his notes on his blog).

### 3.9 Data extraction and synthesis approach

For synthesizing data to answer the MLR's two RQs, we used qualitative coding [159], an approach which we have also applied in several other recent MLR studies, e.g., [67, 71]. We took each qualitative piece of data from each paper, and then performed "open" and "axial coding" [159].

Recall from Section 3.1 that our MLR included two RQs: (1) What are the root causes of low research relevance?; and (2) What ideas have been suggested for improving relevance?. We highlighted texts in each source which addressed each of these two RQs. This ensured full "traceability" of our data extraction and data analysis. We then used the identified phrases to conduct qualitative coding and then grouping the codes, when needed. We show in Figure 8 an example of this process, in which the two aspects (root causes and improvement suggestions) are highlighted in ==pink== and ==green==, respectively. In addition to the above two RQs, we also noticed many interesting quotes in the source which we also extracted and highlighted in yellow in Figure 8.

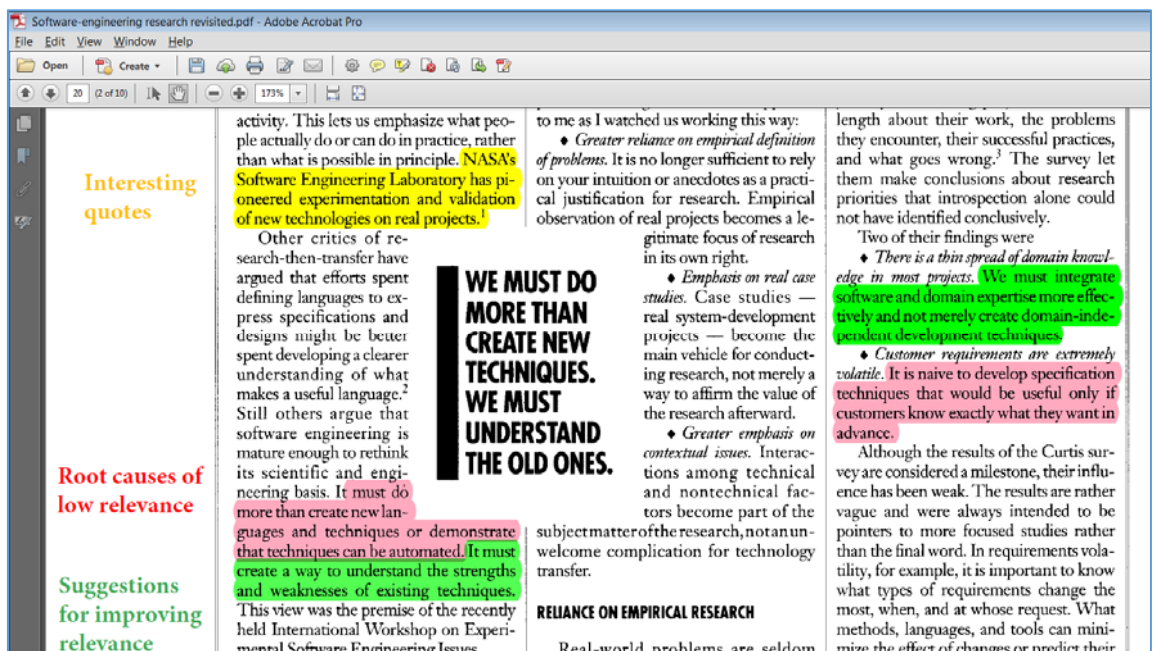

a: Data extraction from the source under study b: Qualitative coding of root causes and improvement suggestions

**Figure 8- An example of the qualitative coding and grouping approach, while ensuring full traceability back to the extracted data in the papers**

Based on the principles of qualitative coding [159], we merged and grouped the identified codes iteratively until we reached a cohesive set of groups based on the extracted qualitative data. For example, see the several groups of root causes of low research relevance in Figure 8, e.g., issues with research approaches (mindset), not considering "context" (approaches too generic). These groups will be discussed in Section 4.1.

Note that each source (paper or GL article) could provide multiple root causes and improvement suggestions and thus we extracted such data and classified each source under multiple root causes and improvement suggestions accordingly. We will describe examples of those throughout Section 4.

In addition to synthesis of individual root causes and improvement suggestions, we planned to conduct two further levels of synthesis to systematically address our two RQs. For each of the root causes (RQ 1) and improvement suggestions (RQ 2), we conducted the first level of synthesis to provide a synthesized summary of the discussions in the literature (to be presented throughout Sections 4 and 5). Our second level of synthesis was to synthesize all issues w.r.t. the root causes and improvement suggestions with the goal of providing "actionable" insights (to be presented in Section 6.2).

## 4 RQ 1: ROOT CAUSES OF LOW RELEVANCE

After extracting the data from all the sources about root causes of low relevance and applying qualitative coding on those data, we classified the results as shown in Figure 9. We have ordered the list of root causes by the frequency of their appearances in the sources under review. Several groups of root causes emerged under the categories "individual researchers" and causes related to collaborations, and thus we have grouped them under those categories. In the following sub-sections, we explain each of the root causes, and provide examples from the sources which have discussed them.

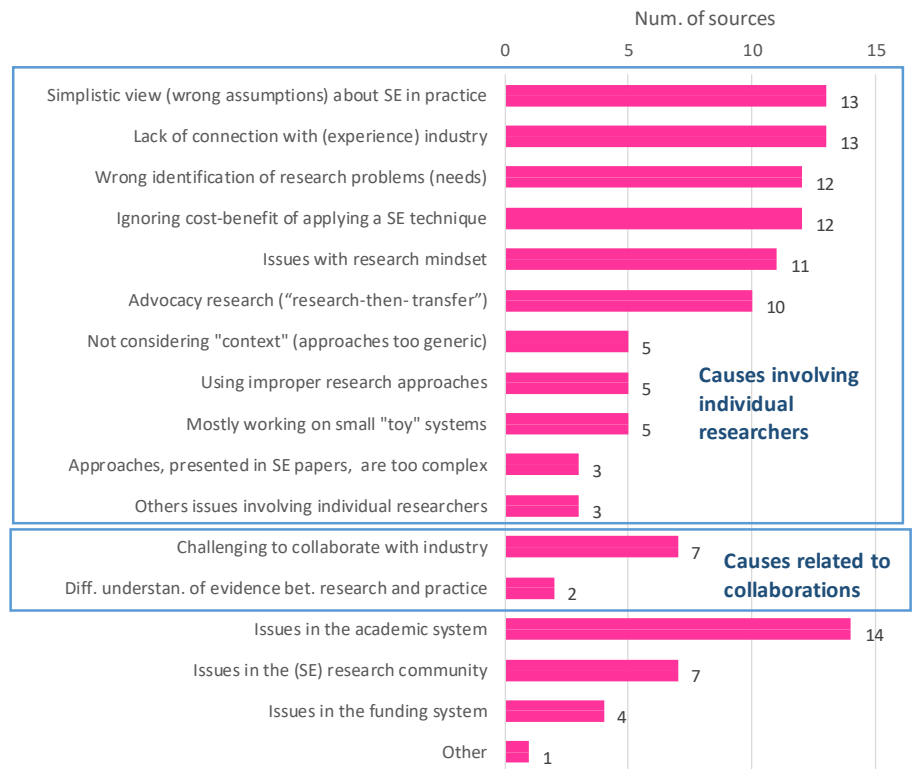

**Figure 9– Grouping of the root causes of low relevance, as discussed in the sources**

## 4.1 Root causes based on individual researchers

By qualitative coding, we classified the issues (root-causes) specific to researchers in several categories, as shown in Figure 9. We discuss those issues in this section.

### 4.1.1 Having simplistic views or even wrong assumptions about SE in practice

Thirteen sources mentioned having simplistic views (wrong assumptions) of SE as an important root cause. Together with "Lack of connection with industry", they were the most frequent under the "researchers" category of root causes.

Focusing on requirements engineering (RE), as a sub-area of SE, S12 mentioned that: "*Too often, research justified as satisfying the needs of industry begins with a wrong or simplified understanding of industry's problems*".

Using data from a survey of 512 practitioners at Microsoft, S15 described a number of reasons given by practitioners to explain why they viewed topics studied in some SE papers as "unwise", including "*questionable assumptions*". Authors of S21 also believed that: "*SE researchers often fall into the trap of making implicit, unrealistic assumptions*".

In an experience report on applying academic software testing results in industry (S6), the author wrote that: "[*Most of the] techniques/tools I tried did either not fully work or were based on unrealistic assumptions*". It then cited a survey paper [160], which based on a survey of 3,785 developers, found that design models (such as those in UML), seem "*rarely used in practice*" and provided the following assessment: "*The use of models in general, and the UML in particular, does not seem to be standard practice and de facto standard in software development, which challenges the assumption on which much of current research is based*".

An IEEE Software paper, entitled "*Practicing what we preach*", S22 also focused on relevance of requirements engineering (RE), and argued that: "*RE practitioners work mainly with documents, so hold your tools. They [practitioners] aren't comfortable with formalizing requirements or using notation to capture requirements*". The argument clearly translates to simplistic views (wrong assumptions) of RE in practice.

In "*Really rethinking formal methods*", S23, David Parnas questioned industrial relevance of formal methods and argued that: "*the models [built and used in formal methods] often oversimplify the problem by ignoring many of the ugly details that are likely to lead to bugs*".

S30 believed that developing SE techniques based on fixed requirements is one root cause of low relevance. S38 also argued that "*[SE] research in a vacuum relies on assumptions that are unlikely to ever match any real context and therefore lead to impact*".

---

<u>Synthesis:</u> Our synthesis of issues mentioned in the sources with respect to this root cause (e.g., the examples above), we can derive the observation that when a researcher has or starts with simplistic views (and often, wrong assumptions) about SE in practice, s/he formulates the research problem and her/his research approaches based on those simplistic views and assumptions. As a result, many of those developed approaches end up having poor applicability and also low relevance in practice (e.g., the above examples). The issue on having simplistic views in research compared to what is out there in the "real" world (SE in practice) can be nicely summarized by the following quote by the well-known scientist, Isaac Asimov: "*Your assumptions are your windows on the world. Scrub them off every once in a while, or the light won't come in*".

---

### 4.1.2 Lack of connection with industry

Lack of connection with industry or experience working in industry was mentioned as another major root cause which appeared in 13 sources. When a SE researcher does not have any experience of working in industry in her/his career, or does not have joint collaborations with industry, it often impacts the way s/he conducts research, and often leads to less relevant research.

S37 mentioned that: "*[a] typical [SE] researcher had never worked in software practice and had no basis for assuming that his or her idea would really work there*" and that "*[most] developers aren't aware of software engineering research*" (again due to lack of connection). S47 also reported that: "*The real problem is that most research academics have little idea what goes on in industry and what research results might be of interest to industry.*"

S21 considered "*no knowledge of existing industry solutions*" as a root cause. S23 argued that "*most software developers perceive formal methods as useless theory that has no connection with what they do*".

S33 saw "*an artificial barrier between industry's problems and research problems*". In a keynote talk (S34) entitled "*Software engineering research under the lamppost*", using students as subjects in empirical studies was seen as a factor weakening connection with industry and thus leading to low relevance. The issue has also been reported widely in other fields, e.g., a paper [121] in the IS community mentioned that: "*While they deal with the same subject matter, practitioners and researchers mix in their own circles, with very little communication between them*".

---

<u>Synthesis:</u> The lack of connection with industry is another main reason behind low relevance, and that such a lack of connection is mostly an issue specific to individual researchers, i.e., an SE researcher has the choice of establishing active linkage with industry or not to. As a cause-effect, this root cause is influenced to some extent by researchers' mindset (preferences) in research (itself discussed in Section 4.1.5).

---

At this point of the paper, we should mention that, due to the complex nature of this subject, there are inherently inter-connections and interdependencies among different root causes, e.g., "lack of connection with industry" could lead to "wrong identification of research problems" (e.g., the above paragraph) and also "ignoring cost-benefit of applying a SE technique". We were aware of this fact during our data extraction and synthesis. However, to prevent our bias in the data synthesis, we decided to record and synthesize the root causes, as they were explicitly reported in the sources (see the example in Figure 8), and did not record nor synthesize interdependencies if they were not explicitly mentioned in the sources. We stayed as close as possible to what the original authors had intended in their writings and prevented our own interpretations (bias) in the synthesis of data.

### 4.1.3 Wrong identification of research problems and needs

Wrong identification of research problems was mentioned as a root cause in 12 sources. S7 assessed whether automated debugging techniques are actually helping programmers. By exploring the technical details of how debugging is done in practice, S6 came to the conclusion that: "*without a clear understanding of how developers use these techniques in practice, the* *potential effectiveness of such [automated debugging] techniques remains unknown*". S13 said that: "*we ignore the real cutting-edge problems that organizations face today*".

S19 stated that: "*practitioners don't seem to want frameworks; they want patterns of context-specific help*". S21 stated that: "*SE researchers lack understanding of industry's true needs*". S28 mentioned that SE papers "*gain little traction with industry because they do not appear to address the practitioner's problems*".

In another classic paper S33, David Parnas stated that: "*The developers' major problems were problems that I had never considered, problems that none of my professors or colleagues thought worthy of discussion*".

Andy Ko, the author of S37, stated that: "*many of our engineering problems simply aren't the problems that software engineering researchers are investigating. Many of the problems in software engineering aren't technical problems, but people problems.*"

Related to the identification of research problems, it would also be timely to mention the following quote from Gordon Clegg's 1969 book entitled "The design of design" [161]: "*Sometimes the problem is to discover what the problem is*".

> <u>Synthesis:</u> Wrong identification of industrial research problems (which industry eagerly needs to be solved) remains a major root cause. As we showed in Figure 1 and discussed in Section 0, when a researcher chooses to work on an "academically-hot" (academically-challenging) topic, while it is possible to conduct high-quality research on such a topic with high rigor, there is a risk that the work will have low relevance by gradually progressing far from industrial "precise" needs. As an example, relevance of the field of formal methods in SE has been questioned by several studies [10, 20, 23].

#### 4.1.4 Ignoring cost-benefit of applying SE techniques

Ignoring cost-benefit of applying SE techniques was mentioned as a root cause in 12 sources. This issue was mentioned in S6, an experience report on applying software testing academic results in industry. The author of the study had tried applying model-based testing, but one problem was that: "*None of the companies*" that he "*worked for in the past five years used any sort of modeling [e.g., using UML]*". He then argued that "*it is important to always state where the models [to be used in model-based testing] come from: are they artificial or did they already exist before the experiments*" and that "*one has to argue and evaluate if the time and effort in developing and maintaining such models for a given system does pay off in the end*".

S10 expressed that the utility of formal methods "*is severely limited by the cost of applying the techniques*". S20 also reported that "*introduction of formal methods consumes considerable resources*". Another paper on relevance of formal methods, S23, stated that: "*Even on small examples, it often appears that the models [used in formal methods] are more complex than the code*". Complexity in this context related to cost-benefit of applying the technique. S24 also stated that: "*formal notations and analysis techniques are [often] hard to learn, provide no clear benefits and require too much effort*". S41 also reported that "*practitioners saw more cost than benefit in its [formal verification] use*".

S13 mentioned that "*research solutions offer too small improvements in relation to risk*". S15 found that "*practitioners deemed the cost of using and maintaining a particular tool to be higher than the benefit gained by using the tool*".

S21 also argued that "*an engineering solution is useful if the benefit gained is significantly greater than the cost*". The study went further to state that: "*There is little discussion in [SE] research of the full costs of a technique (including acquisition, training, and disruption to existing processes). There are seldom arguments made as to why the benefits are significantly greater than the costs and how this technique will reduce project risk*".

S28 stated an even harsher opinion: "*Rather than showing a better, more efficient way to do things, they [most SE papers] call for additional work that has no obvious benefit*".

> <u>Synthesis:</u> When SE researchers propose new approaches and techniques, they often fail to conduct/report either a qualitative or quantitative cost–benefit analysis concerning their application. In such a case, when practitioners read about those techniques in the paper, they often wonder how much benefit they would get from applying such techniques and how much cost applying them would incur. It is quite sad that many SE papers are not really considering the efforts (in most cases non-trivial) for generating the inputs (e.g., models) needed for applying a technique. Without clear and precise answers to these questions, chances of industrial applications are often slim.

### 4.1.5 Issues related to researchers' mindset

Eleven sources mentioned root causes related to researchers' "mindset" (preferences). In "*Making research more relevant while not diminishing its rigor*" S17, Robert Glass argued that "*most researchers strive for rigor and see relevance as a lesser goal*", thus calling researchers to have a mindset shift to also value relevance, while focusing on rigor.

S22 argued that: "*While researchers tend to address challenges that they believe have the potential to be transformative in nature, practitioners are more often intent on solving immediate practical problems*", which was another need for mindset (preferences) shift. In a classic paper S30, Colin Potts argued that: "*It [SE] must do more than create new languages and techniques or demonstrate that techniques can be automated*". The tendency to create new languages and technique is often related to researchers' mindset and "school of thought".

David Parnas shared in S33 his story of spending time outside academia and working with software developers at Philips Corporation in the Netherlands. He mentioned that as he "*watched their [developers'] work, and tried to understand their discussions*", he realised that his "*[research] vision (which I [he] later discovered was shared by several other researchers) was completely wrong*".

In another classic paper S41, Robert Glass believed that the SE research was in "crisis", due to lack of relevance. One main reason he said was "*not using the scientific method*", which had to start with "*observing the world*", and also not analyzing critically if a new model or approach is really needed.

Researchers' tendency to have strong opinions in their mindset, which is often one main root cause for low research relevance since they tend to put more emphasis on rigor, have been discussed widely in science in general, for example, an article in Washington Post [162] put the issue boldly as: "Academia is a cult". The issue of research relevance has also recently been discussed among some SE researchers in Social media, e.g., a senior SE researcher said recently on Twitter[1]: "*There is little difference between academic communities and religious cults. They defend a set of beliefs, at any cost, against external influences*".

Synthesis: All the sources which mentioned this root cause expressed the fact that the way researchers think about SE research, as well as their preferences for topics that should be studied, has a profound impact on the relevance of their research activities.

### 4.1.6 Conducting "advocacy" research

Conducting "advocacy" research was mentioned as a root cause in 10 sources. Robert Glass (S41) characterized advocacy research as: "*(1) conceive an [research] idea, (2) analyze the idea, and [then] (3) advocate the idea*".

Colin Potts (S30) referred to advocacy research as the "*research-then-transfer*" approach. Potts added that: "*Unfortunately, software engineering research is often more solution-driven than problem-focused*". S9 mentioned that: "*Engineering research must be problem-driven, account for real-world requirements and constraints*". S23 mentioned: "*We need research, not advocacy: When we find that people are not adopting our methods, it is tempting to try "technology transfer" and other forms of advocacy*". S54 mentioned that "*Solution-oriented papers that propose [generic] methods or process models are viewed as too general [by practitioners]*".

S42 categorized the SE research into two incommensurable paradigms: (1) The rational paradigm: which emphasizes problem solving, planning and methods; and (2) The empirical paradigm: which emphasizes problem framing, improvisation and practices. The paper argued that "*the Rational and Empirical Paradigms are disparate constellations of beliefs about how software should be [developed]*" versus how "*it is created [developed]*". The rational paradigm looks similar to advocacy research. S42 went on to state that: "*This [having two very different SE paradigms: rational versus empirical] creates enormous confusion and conflict in the SE community*" and "*Having two paradigms in the same academic community causes miscommunication [4], which undermines consensus and hinders scientific progress*".

We can characterize and better understand advocacy research by using the model proposed in S31, which conceptualized the collaboration modes (styles), or "degree of closeness" between industry and academia, as shown in Figure 2 (discussed in Section 2.1.5). There are five levels in that model, which can also be seen as maturity levels: (1) not in touch, (2) hearsay, (3) sales pitch, (4): offline, and (5) [working as] one team.

---

[1] https://twitter.com/lionel_c_briand/status/1090878723732242432

We argue that, in levels 1-3, there is really no IAC, since researchers working in those levels only vaguely identify a general challenge of the industry, develop a solution for it, and then (if operating in level 3) approach the industry to try to convince industry to adopt the solution (advocacy research). However, since such techniques are often developed with a lot of simplifications and assumption often not valid in industry [109, 113], they fail to be applicable (work) in industry, thus have low relevance. IAC which would increase research relevance really occurs in levels 4 and 5. Of course, level 5 is the most ideal case in which both parties work as "one team" to identify a specific challenge. They then iteratively evaluate and validate draft solution approaches and finally (and hopefully) will deploy a tailored solution in the context of the industrial partner. Last, but not least, we want to emphasize that it is the opinion of many researchers and practitioners [163] that "technology transfer" (only from academia to industry) is not a suitable term for IAC, but better terms are "(mutual) knowledge transfer" and "knowledge exchange".

> **Synthesis:** Advocacy research seems to be another major factor leading to the relevance problem. When researchers follow such a research paradigm, in which the research work is mostly done far from "practice" (the real world) and the idea is to "advocate" research only after it is done, the chances of applying the final approach in practice would be marginal.

### 4.1.7 Research focus being too generic and not considering context

Not considering "context" was mentioned as a root cause in five sources. A 2018 paper, S29, defined context in SE as the factors "… *within which SE methodologies and technologies are to be applied and therefore the context that research needs to account for, if it is to be impactful*". S29 classified contextual factors into three categories: human, domain and organizational factors. The paper believed that "*most of the research is insufficiently grounded in real development contexts*". Authors of S38 also stated that there is a "*shortage of context-driven research*" in SE. S21 also argued that: "*some SE researchers invent solutions without considering whether they can be used in a real--world context*". S5 argued that: "*It is not enough to develop new theories and provide evidence. The practitioners who are the audience for our evidence must be able to understand our theories and findings in the context of their work and values*".

Related to the issue of context, many sources believed that focus of most of SE research papers has been too generic (general) and thus have not considered context. For example, S30 mentioned that "*Much of the research-then-transfer [SE research] work that seeks to develop completely general tools is naïve*".

> **Synthesis:** Paying attention to context (human, domain and organizational factors) seems to have been overlooked to a large extent in the SE literature so far. While a small ratio of papers has considered contextual factors (e.g., approaches for testing mobile applications), many other SE research efforts have focused on "generic" approaches.

### 4.1.8 Using improper research approaches

Five sources reported that improper research approaches, which are established in the research community, lead to low relevance. In a blog post (S8), an SE researcher argued that "*Studying human systems like software engineering organizations requires research method skills that cannot be found easily among [SE] researchers*".

Using data from a survey of practitioners in Microsoft, S15 provided two reasons given by practitioners to explain why they viewed topics studied in some SE papers as "unwise": (1) empirical study being "non-actionable", and (2) generalizability issues in most empirical studies. These two shortcomings clearly relate to weaknesses of research approaches established in the SE community. The SE community as a whole has gained some maturity in the latter issue since most (well-written) empirical studies now discuss generalizability issues to some extent. However, since most SE papers and empirical studies do not provide concrete actionable insights for practitioners, practitioners are often puzzled what to do or take-away after reading a SE paper.

S38 mentioned that: "*A common misperception [in the community] is that long-term, high-risk research (sometimes called basic research) must [could] be free of practical considerations*".

In S41, Robert Glass said that, due to mostly being rooted in CS and mathematics, most people in the SE research community use the "analytical" research method which consists of the following steps: Propose a formal theory or set of axioms, develop a theory, derive results, and if possible compare with empirical observations.

> **Synthesis:** The examples discussed above show that traditional research approaches, which are established in the SE research community (often rooted in CS and mathematics), may lead to low relevance.

### 4.1.9 Working on small ("toy") systems and not considering scalability

Working on small "toy" systems was mentioned as a root cause in five sources. To assess whether automated debugging techniques are *actually* helping programmers, S7 critically studied a number of empirical studies in this area, and found that: "*most programs studied [in empirical studies] are trivial, often involving less than 100 lines of code*". S13 mentioned that "*we [SE researchers] spend too much time with toy problems that do not scale, and as a result, have little applicability in real and complex software projects*".

S29 argued that most of SE papers propose "*… solutions that are not applicable or scalable, even with additional engineering effort*". S37 also believed that: "*One of the most important issues in SE is to define what will and what will not scale*".

> **Synthesis:** When assessing research approaches on small ("toy") software systems, most of the issues (regarding scale and magnitude) of dealing with real-world large-scale systems do not emerge. Thus, the approaches developed in such research efforts will hardly be applicable in industry. Of course, the issues of having "access" to large-scale software systems for research is an issue here, which could be addressed by joint IAC. Many SE researchers conduct research on open-source systems and repositories as a way to work on large-scale systems.

### 4.1.10 Approaches, presented in SE papers, are too complex

Three sources mentioned that many researchers develop approaches which are too complex. In a paper entitled "*What industry needs from architectural languages*", S51, the authors believed that: "*Heavyweight and complex architectural languages often deter practitioners*". S19 framed the issue nicely as follows: "*complexity impresses, but prevents impact*", in which he meant that, the higher complexity of a solution (SE approach), the lower its applicability in general.

In criticizing the complexity of formal methods, S10 mentioned that "*The mathematical abstractions embodied in notations such as Z and Timed CCS facilitate brevity and precision, but they do not necessarily contribute to clarity*". Clarity of a solution (SE approach) is an aspect which could directly impact its applicability. On the other hand, the software industry strives for simplicity. For example, GitHub's Director of Systems told the crowd in an industry conference that "*the GitHub team tends to shy away from unnecessary complexity in the systems that it builds*", and put forward "*the argument for simplicity in software development*" (https://www.youtube.com/watch?v=iXkJh0629RE).

Even in technical tasks such as architecture, engineers target simplicity, e.g., a practitioner argues in a blog-post [164] about: "*Why simple is so complex: the complexity of simplicity in software engineering*". In pursuit of simplicity, some software practitioners even have adopted a principle named *Occam's razor* [165]. It is the problem-solving principle that states that "*simpler solutions are more likely to be correct than complex ones*". When presented with competing hypotheses to solve a problem, the principle recommend one to select the solution with the fewest assumptions [165].

> **Synthesis:** While practitioners normally strive for simplicity, researchers often strive for complexity in their proposed approaches, which is often being the norm of the academic community – let us recall that "*complexity impresses, but prevents impact*" (S19). Such a mismatch between research and practice is leading to decreased research relevance.

### 4.1.11 Other root causes specific to individual researchers

When classifying the root causes specific to individual researchers, we also notice a few "other" root causes in the sources which only appeared once and thus we did not create separate categories for them. We report three such sources in the following paragraph.

S20 mentioned that "*[many] practitioners state that they consider our topic [formal methods] too theoretical and not worth their time*". In criticizing formal methods, it also stated that: "*Formal SE approaches are subject to the same errors as programs*". As another root cause, S21 mentioned that "*Often the only users of the technique are the researchers themselves, and they have made scant attempt to empirically evaluate the technique by practitioners*". S44 was entitled "*Towards improved Adoption: effectiveness of research tools in the real world*". The authors mentioned that: "*One reason for this limited adoption is that researchers often fail to provide real-world evidence that the methods and tools they develop are of potential value to industry practitioners*".

## 4.2 Root causes originating from industry-academia collaborations

During our qualitative coding, another group emerged which was in the scope of IAC [106]. When coding those root causes, we classified them into two categories:

(1) Challenging to collaborate with industry
(2) Different understandings of evidence between researchers and practitioners

### 4.2.1 Challenging to collaborate with industry

Seven sources mentioned that it is, in general, challenging to collaborate with industry and that this hurts research relevance. For example, S8 reported that "*Working with industry requires time: You have to build up a reputation, gain trust, and may still not be able to get the data you need*". S22 mentioned that "*People from industry seem reluctant to share data with researchers owing to confidentiality agreements they must respect*".

Moreover, S35 mentioned that: "*it [IAC] is expensive, takes a lot of time and doesn't generate many papers*". The same issue was phrased in S41 as: "*Software research that involved evaluation in realistic setting is expensive in terms of both time and money*".

The challenge of collaborations with industry has also been discussed in other fields, e.g., in psychology [88]: "*Practitioners and researchers have often held stereotypical views of each other, with practitioners viewing researchers as interested only in methodological rigor whilst failing to concern themselves with anything in the real world, and researchers damning practitioners for embracing the latest fads, regardless of theory or evidence*".

> **Synthesis:** Since it is challenging to collaborate with industry and it requires (extensive) time and effort, many researchers conduct their research without any industry collaborations, which in turn, could lead to decreased research relevance. However, the benefit of conducting research via IAC is that it could address many of the previously-listed root causes, e.g., wrong identification of research problems, or ignoring cost-benefit issues of applying SE techniques.

### 4.2.2 Different understandings of evidence between researchers and practitioners

Two sources mentioned this root cause. S5 argued that "*Practitioners and researchers have different ideas about what makes good evidence*". S5 argued that practitioners prefer to look at field or case studies performed in context to decide if the study's environment is similar to their own. In contrast, researchers prefer to use controlled experiments, simulations, and other more general studies. As a result: "*… without the tie to what they perceive as a 'real' situation, practitioners often ignore the results of these more general studies [research papers]*".

S46 argued that: "*Practitioners, being trained with an engineering mindset, expect generalized and quantitative results. They want to know what by percentage productivity will improve if they adopt a new practice, but this is a level of precision that no honest scientist can offer them today*". We can observe that most SE papers have not provided quantitative measures on productivity improvements (in software teams) when a given SE approach is used.

> **Synthesis:** Since understanding of proper evidence between researchers and practitioners are sometimes different, this negatively impacts relevance of SE papers, when they are read by practitioners.

## 4.3 Root causes in the academic system

Researchers are mostly working in the academic institutions and thus should follow the norms and expectations of the typical academic system, e.g., number of publications and citations often count more than IAC, or industrial relevance or impact. Fourteen sources mentioned root causes related to the academic system.

S6 raised the issue of "*Being rewarded for number of publications, in contrast to other engineering fields that put more focus on patents and industry collaborations*". S7 mentioned that: "*Quantity [of papers] counts, not how interesting or novel or relevant the work is*". Incentive misalignment as a cause for the relevance problem was also mentioned in S21, S25, S28, S35, S46 and S48.

S26 phrased the issue as the "*publication numbers game*". There are debates about this issue in other fields too, e.g., "*Publications in peer-reviewed journals continue to be the key performance indicator within academia: whether anyone reads them is a secondary consideration*" [166].

S9 argued that "*academia does not value … [industrial] impact*". It further mentioned that "*the paradigm of research in engineering is somewhat different from that in natural sciences or applied mathematics*" and "*No wonder many software engineering researchers find it difficult to perform high-impact research – they have to comply with computer science expectations in terms of research contributions*". S21 argued that, to meet expectations of the academic system, "*professors are interested in publishing papers only.*"

S25 talked about "*disincentives to leave the university for periods of time to work in industry*", "*emphasis on more theoretical journals for 'credit' toward tenure and/ or promotion*", and "*a learning environment in which (unlike medical school) students are taught by professors with little or no experience in the 'real world'*".

S37 argued that "*many CS departments don't view social science on SE as computer-science research*". S41 argued that "*Those [researchers] using empirical methods were not admired by their more 'traditional' colleagues*" and that "*'Pure' (basic) research has more academic 'respectability'*". S48 mentioned these issues: "*Our CS legacy (emancipating ourselves as an engineering discipline)*" and "*counting papers*".

> Synthesis: There are various issues and legacies in the academic system, e.g., the incentive system and the CS roots of SE, often mostly outside the control of an individual SE researcher, which intensify the relevance problem in one way or another.

## 4.4 Root causes in the (SE) research community

Seven sources mention issues related to the (SE) research community. S3 proposed a "*Manifesto for applied requirements engineering research*", an item of which was: "*collaboration and transparency [in the community] over competition and protection*".

S17 claimed that "*Academic journals [in SE] don't reward it [relevant research]*". Similarly, S30 mentioned that "*Technology transfer is undervalued [in the community]*". S33 raised the issue by stating that "*referees judge papers by the standards of non-engineering fields [e.g., CS]*". The challenge of publishing applied ("actionable") research has been stated like a "paradox" [119] in other fields too, e.g., a paper [119] in management science mentioned that: "*The 'relevance literature' often moans that the publications of top-ranked academic journals are hardly relevant to managers, while actionable research struggles to get published*".

S38 argued that "*we must distance ourselves from the research tradition in computer science that emphasizes proofs of correctness and completeness instead of practicality and scalability, and that tends to believe in universal concepts and solutions*", which denotes another symptom of community "culture", negatively impacting research relevance. With a similar argument, S9 mentioned that: "*We have to admit that in most computer science departments, to which most software engineering researchers belong, this [valuing research impact] isn't the case*".

S48 mentioned that "*SE research is more driven by 'fashion' than needs, a quest for silver bullets. We can only blame ourselves*". It also argued that papers in industry tracks of most SE conferences are not "*seen as first-class citizens*".

S45 mentioned that: "*The observation that practitioners rarely read the academic literature leads us to consider the wider question of whether academia cares enough about the practical relevance of their output to change their dissemination patterns*". We can see that "academia" here refers to the SE community.

Similar challenges have been discussed in other fields too. For example, a paper [121] in the IS community argued that: "*Research communities are very powerful, because they control both the supply and demand of IS research*". The key to getting published in a field of science, and hence academic "survival", is to find a "research community" that is interested in one's work. In such an environment, the real audience for research becomes other researchers on journal editorial review boards and conference program committees (which rarely include practitioners). These people dictate the "demand" for research — what is considered important or high quality research and, therefore, what gets published.

We argue that the situation is quite similar in SE as well. The readership of most leading SE journals and conferences is mostly academics. This means that, similar to IS, the SE research is written mostly by academics, reviewed by academics, and read mostly by other academics. As Keen [87] succinctly puts it: "*IS research is in danger of talking about itself to itself*" [121]. One could wonder whether this issue is also largely applicable in SE.

Research has also been done the "sociology" of scientific communities, e.g., [167], outside the SE community. These studies have investigated various social aspects of scientific communities such as groupthink, group opinions and group mindsets, and how these aspects could impact the forming of norms, opinions, and beliefs in community. It would be interesting to replicate some of the studies in the SE community.

Let us mention that, as discussed in the above paragraph, there is the concept of "group mindset" [167] as well, in addition to the individual researchers' mindsets. However, this time, it is on the group level. This concept is also often called the "(dominant) group opinion" [168]. As per sociology and psychology literatures, individual mindsets may be partially influences by (dominant) group opinions, but only to some degree [168]. Thus, they are not always the same.

> Synthesis: As we see from the above examples, in addition to the categories that have emerged in the previous sub-sections, several other root causes specific to the (SE) research community have also been reported. One example is the readership of academic papers, i.e., they may be written to please reviewers rather than to share results with practitioners.

### 4.5 Root causes in the funding system

Four sources mentioned issues related to the academic funding system. S12 discussed about general "*lack of funding*" for relevant and practical research. S45 argued that "*researchers are neither incentivized nor funded to make their work relevant and practical*". S35 argued along the same lines.

S39 also mentioned that there is no funding for building and maintaining large software systems in academia that could be used for relevant research. It explained this funding-related challenge as follows: "*in computer science or any part of it (and software engineering is generally viewed as a subset of computer science), the idea that we would actually do something, rather than talk about someone else's artifacts, is alien to the funding process*".

In this outset, we should mention that, funding systems in certain jurisdictions have realized the importance of relevant research and are actually "encouraging" such research by requiring to have both industry and academia in project proposals, e.g., the Horizon 2020 funding system of the European Union, the "Engage"[1] and Collaborative Research and Development (CRD)[2] grant programs in Canada. Thus, the root causes in the funding system as discussed in four sources are often local issues (specific to certain countries).

> Synthesis: We can see that even due to certain factors in the funding system, SE researchers in certain countries are sometimes discouraged from conducting relevant research.

### 4.6 Other root causes of low relevance

One source mentioned another root cause of low relevance. S13 argued that "*Software research is biased toward huge projects*" and, thus, small to medium size projects may not benefit from most SE papers.

## 5 RQ 2: SUGGESTIONS FOR IMPROVING RESEARCH RELEVANCE

Similar to what we did in the case of RQ 1, we used the qualitative coding approach [159] to synthesize and group the suggestions for improving research relevance. Figure 10 shows the grouping and frequency of suggestions for improving relevance, as discussed in the sources. Similar to Figure 9, several of the categories relate to improvements that should be done by individual researchers and, thus, we have grouped them as such.

As we noted in Section 4.1.2 regarding the root causes, due to the complex nature of this subject, there could be interdependencies among different improvement suggestions. For example, although we have placed "Need for using proper research approaches" under improvement suggestions involving researchers, choice of research approaches could also be influenced by funding agencies as they review project plans and may favor certain approaches over others. However, we should again note that, to prevent our bias in data synthesis, we recorded and synthesized the improvement suggestions, as they were explicitly reported in the sources, minimizing our own interpretations. Thus, if a source had mentioned an improvement suggestion involving researchers, we recorded that data as such.

---

[1] http://www.nserc-crsng.gc.ca/Professors-Professeurs/RPP-PP/Engage-Engagement_eng.asp
[2] http://www.nserc-crsng.gc.ca/Professors-Professeurs/RPP-PP/CRD-RDC_eng.asp

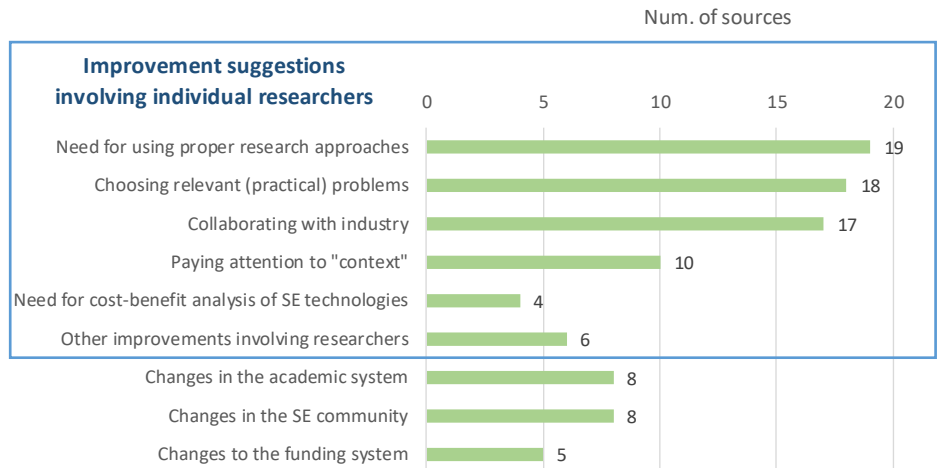

**Figure 10- Grouping of suggestions for improving research relevance, as discussed in the sources**

As expected, most of the suggestions for improving relevance, discussed in the sources, correspond to the root causes (Section 4) and in most cases, they are in fact suggested solutions to those root causes, e.g., S41 suggested adopting the "Industry-as-laboratory" approach to replace the "Research-then-transfer" approach.

To see the relationships between the root causes and the improvement suggestions more clearly, we "mapped" their relationships as a cause-solution graph in Figure 11. Each improvement suggestion is connected to an item in the "causes" category via an edge labeled "Will address", as discussed in the sources under review (note that, again, we minimized our own interpretations, to prevent bias). The order of the items in the two sides in Figure 11 are the same as the orders in Figure 9 and Figure 10, which are based on the frequencies of each item, discussed in the sources and as synthesized in our qualitative coding.

Figure 11 illustrates that none of the sources provided improvement suggestions for a few of the causes, e.g., for "Working on small/toy systems". It could be the case since such causes are seen as rather "trivial" and thus authors of the sources did not bother to provide explicit suggestions for improvement of research relevance with respect to those causes. For this cause, the obvious suggestion would be to work on large-scale systems rather than on small/toy systems.

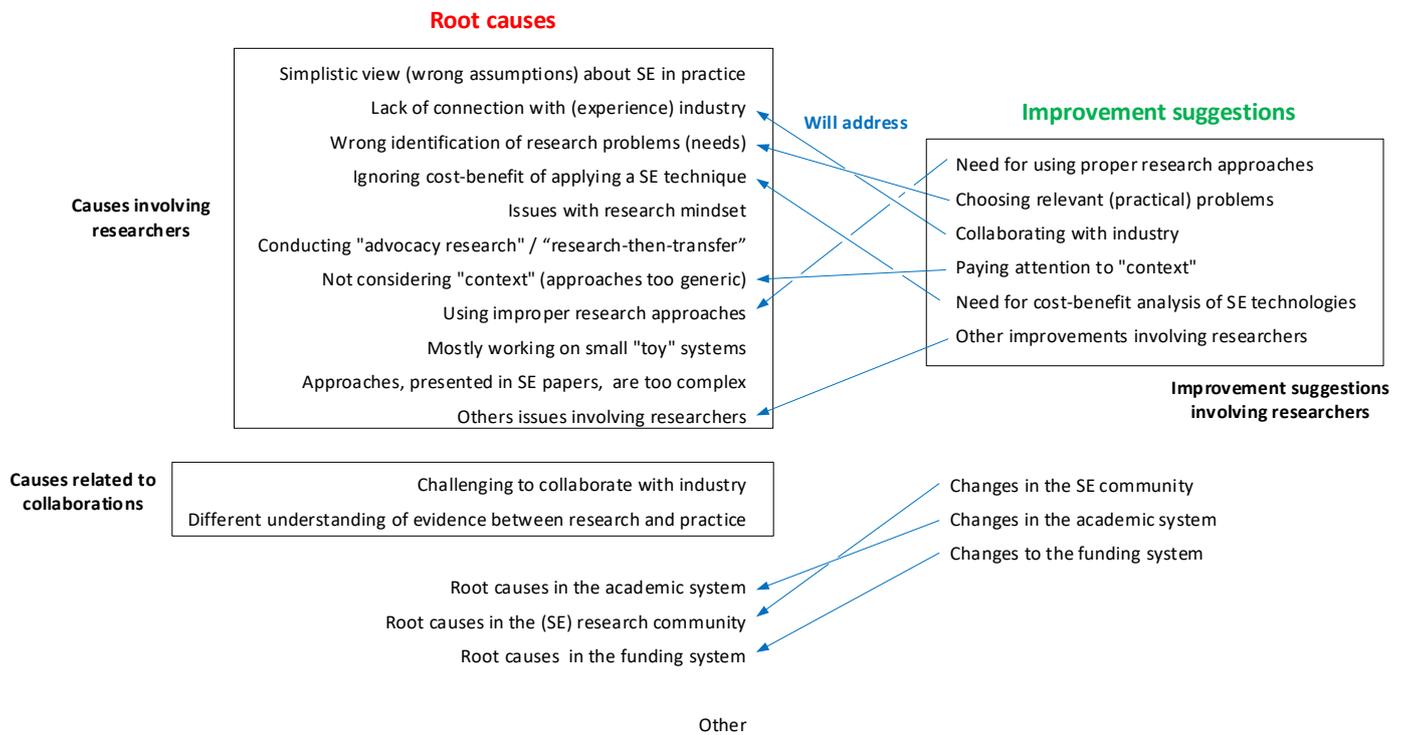

**Figure 11- Mapping the relationships between the root causes and the improvement suggestions**

When analyzing the improvement suggestions, another important issue worth looking into, was the "proposed mechanisms" for improvements, as discussed by the authors of the analyzed sources. We believe such mechanisms could be as interesting and important as the suggestions themselves, because "how" to improve is a precondition for actually improving the root cause behind low relevance. However, in almost all of the cases (as presented below), we found that the authors of the sources had only discussed the improvement suggestions and not provided the how-to's for making those improvements. To shed some light on the how-to aspect, several papers in the literature have provided some ideas, e.g., for identification of relevant SE research problems, a few approaches (mechanisms) have been published, e.g., [52, 169].

To clarify this issue, we visualize the relationships between root causes of low relevance, improvement suggestions, and improvement mechanisms in Figure 12. It should be recalled once again that the focus of this paper is on the first two issues (root causes and improvement suggestions), since widening the scope to improvement mechanisms would lead to an impractical broadening of the study and to the inclusion of further papers (e.g., [52, 169]) and GL sources. This extension can be done in a follow-up study. However, to still provide some insights, we provide some references to selected literature about improvement mechanisms throughout the next sub-sections.

We can use the above example again to discuss the relationships in Figure 12. While "Wrong identification of research problems" is a root case, "Identification of relevant SE research problems" can "conceptually" address the root case. To "concretely" do that, the approaches (mechanisms) proposed in [52, 169] could be used. For example, a recent study [169] explored what industry *wants* from academia in software testing by gathering practitioners' opinions.

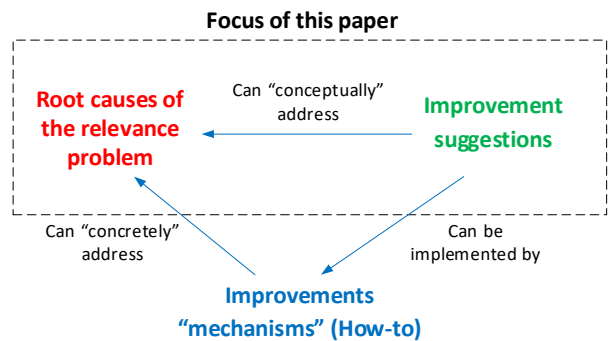

**Figure 12- Relationships between root causes of the relevance problem, improvement suggestions, and improvement mechanisms**

## 5.1 Improvement suggestions involving individual researchers

The improvement suggestions, involving individual researchers, were classified into six groups as reported next.

### 5.1.1 Need for paradigm shift in research approaches

Nineteen sources called for using proper research approaches and a paradigm shift in research. S2 suggested adopting research methods that are "*contributing to relevance*" as listed below: action research, case study, field study, descriptive/exploratory survey and interview studies.

S5 mentioned the "*need for [research strategies to deal with the imperfect knowledge and uncertainty [in SE]*". S6 raised the "*need for research that take into account the way programmers actually debug*". S13 suggested to researchers to "*propose lightweight techniques that are low-risk*". S19 suggested to "*Keep things simple: 'real' as in 'real world', not 'real papers'*", and to "*make things fit into real processes: complexity impresses, but prevents impact*".

S23 argued that: "*We need research, not advocacy*" and that: "*More [grant] money won't help; more thinking will*". It also mentioned that: "*When we find that people are not adopting our methods, it is tempting to try "technology transfer" and other forms of advocacy. When that fails, which it has, we must return to research and look seriously for ways to improve those methods. It is our job to improve these methods, not sell them. Good methods, properly explained, sell themselves*".

S23 called for the need to "*rethink the research direction in RE*". S27 called for "*bottom-up research and a succession of case studies*". S30 mentioned the "*Need to understand the strengths and weaknesses of existing techniques*" and also the "*Need for approaches on evolutionary requirements*".

S34 called for "*a paradigm shift ... from purely theoretical and building-oriented to experimental*". S36 suggested to researchers to "*see the world [SE practice] as it is*". S36 suggested "*building and maintaining large software*". S41 suggested using the "*industry-as-laboratory*" approach, in which "*researchers identify problems through close involvement with industrial projects, and create and evaluate solutions in an almost indivisible research activity*" (S30). S48 advocated for "*inductive*" research, i.e., "*working from specific observations in real settings to broader generalizations and theories*".

As reported in Section 4.1.6, S42 categorized the SE research into two incommensurable paradigms: (1) The rational paradigm: which emphasizes problem solving, planning and methods; and (2) The empirical paradigm: which emphasizes problem framing, improvisation and practices. S42 argued that these two paradigms correspond, roughly, to the basic and applied research in SE. S42 made two bold claims: "*The rational paradigm undermines the credibility of the SE research community*" and "*Embracing the Empirical Paradigm is crucial for retaining scientific legitimacy, solving numerous practical problems and improving SE education*".

In Section 3 we separated the root causes of "Issues related to researchers' mindset (preferences)" from "Using improper research approaches", although they had some conceptual similarities. But in the case of improvement suggestions, discussions in the paper morphed improvements in both of these issues together and thus we grouped them together for the case of improvement suggestions (the above discussions).

Synthesis: To increase relevance, most SE researchers have to first individually accept the need for a paradigm shift in research approaches, from purely theoretical to experimental by more usage of approaches such as: action research, case study, and field study; need to understand the strengths and weaknesses of existing techniques; need for research strategies to deal with the imperfect knowledge and uncertainty in SE; need for lightweight techniques (most techniques in SE papers are quite heavy-weight); and to make things fit into real-world processes.

Improvement mechanisms: Few "mechanisms" were reported in the sources. However, it is evident that a paradigm shift in research approaches is possible for both individual researchers and the SE community level. Perhaps, insights from "sociology" of scientific communities, e.g., [167], should be incorporated for making such initiatives.

### 5.1.2 Choosing relevant research problems

Choosing and working on relevant problems would obviously contribute to research relevance. Eighteen sources mentioned this suggestion. S18, an IEEE Computer paper entitled "*Making SE research relevant*", suggested to researchers to: "*Ensure research is relevant, reflects the needs of practice, and can confidently declare the reason for conducting it*". Selecting the "right" topics for IAC is sometimes easier said than done, but there are guidelines for it, e.g., [169].

S9 suggested to researchers to "*understand the problems and priorities of the many industry sectors that develop software*". S13 suggested to "*focus on solving real problems the industry has*". S15 suggested to "*understand and use the real problem as a beacon*". S22 suggested to "*investigate the problems to identify unanswered research questions, [and to] conduct a more informed research*". It also argued that "*researchers who listen to customers' problems are better able to add value to businesses by understanding problems*".

S24 argued that: "*If requirements engineering researchers would follow the first rule of any requirements engineer, i.e., 'Know thy customer', more of the research would prove to be helpful in practice*". S27 suggested to "*work on high-priority, well-defined industrial problems*". S29 argued that "*we must understand how the research relates to industry (I would argue that we do not)*". S31 suggested to "*distinguish important problems from wishful thinking*". S34 suggested to "*Present [do] more relevant research. Show the work that changes how engineers do their job*".

Synthesis: Choosing relevant (practical) research problems is perhaps among the most "concrete" steps for making SE research more relevant. The higher the relevance of a research topic (paper), the higher likelihood that the end result (paper) will be relevant.

Improvement mechanisms: Apart from our study pool, i.e., papers that discussed "relevance", some recent work has been reported in SE for identification of research problems (needs) which could be relevant. For example, [52] explored what industry wants from academia in software testing by gathering practitioners' opinions. An experience report for selecting the "right" topics (ensuring both rigor and relevance) in IAC in software testing was reported in [169]. Misirli et al. [170] proposed four fitness criteria as guidelines for topic selection in industry experiments: (1) concreteness of the topic, (2) suitability of the topic for experimentation, (3) relevance of the topic to research community, and (4) prior domain experience (of researchers).

### 5.1.3 Collaborating with industry

Collaborating with industry is an obvious way to increase (chances of) research relevance. Seventeen sources mentioned this suggestion.

S16 considered "*collaboration [with industry] as an enabler*". S11 also suggested to: "*Connect with software engineers*". S18 suggested to "*work with industry to find problems and frame your work, [but] do not work with industry to solve (their) concrete problems*", and "*having skin in the game*" (i.e., to be engaged in industrial projects). S13 suggested that "*we should be visiting industry conferences way more often. When a researcher ventures into an industry conference with interesting data, it does seem to generate excitement and good discussions, at the very least.*"

S24 advocated "*two-way [technology] transfer*", i.e., "*knowledge creation and exchange between researchers and practitioners*". "*Successful technology transfer will depend on continuing two-way collaborations between researchers and practitioners. They need one another due to their complementary skills, knowledge and experience*". S29 argued that: "*Collaborative research between industry and academia is a requirement for context-driven research with significant impact*".

In "*Successful software engineering research*" (P33), Parnas mentioned that: "*I would never have realised the nature of the problem, unless I had been working on that project, reviewing development documents, and sitting at that lunch table*". A similar recommendation was made in P47: "*The solution to the problem of industrial research relevance is for academics to be willing to move outside the university bubble, to go out and interact with people in industry*".

---

Synthesis: Many researchers and practitioners are advocating for more IAC. Although it is understandably challenging to collaborate with industry (see Section 4.2.1), and as per other relevant literature [106, 110, 111], "*where there's a will, there's a way*". Such a "will" often comes from the mindset, awareness and appreciation of an SE researcher for research relevance.

Improvement mechanisms: A recent Systematic Literature Review (SLR) of challenges, patterns (best practices, i.e., what to do to ensure success), and anti-patterns (what not to do) in IAC was reported in [106], which was based on review of 33 primary studies, e.g., [110]. According to the SLR results [106], some of the best practices for establishing and maintaining IAC are: (1) Ensure engagement and manage commitment; (2) Consider and understand industry's needs, challenges, goals, and problems; (3) Ensure giving explicit industry benefits and solve the right problem; (4) Be Agile; and (5) Have mutual respect, understanding, and appreciation.

Using the results of that SLR [106], a follow-up empirical study was reported in [111] which assessed the extent of those challenges, patterns and anti-patterns in a dataset including 47 opinion data points, corresponding to multiple IAC projects conducted in 10 different countries.

---

### 5.1.4 Paying more attention to context in research

Ten sources called for paying (more) attention to context when conducting research. S29 classified contextual factors into: human, domain and organizational factors.

S16 believed that "*SE approaches should consider humans [human factors]*". S18 suggested to "*present patterns of context-specific help reflects the needs of practice*". S29 suggested to pay "*attention to human factors*" and to "*foster context-driven research*" in SE research. S38 also highlighted the need for conducting context-driven research which it referred to as "bottom-up research". S48 also believed that "*context factors matter a great deal*".

---

Synthesis: In response to the root cause of "Not considering context (research focus being too generic)", in Section 4.1.7, members of the community have advocated to consider context when conducting SE research. Outside our study pool, in SE in general, the issue of (research) "context" has also received more attention lately. For example, Petersen and Wohlin argued [171] that: "*in order to draw valid conclusions when aggregating evidence [in industrial studies], it is important to describe the context in which industrial studies were conducted*". That paper [171] structured the context for industrial studies in SE by breaking it into six different context facets: product, processes, practices and techniques, people, organization, and market. The paper also provided a checklist [171]. The aim of the study was to aid researchers in making informed decisions concerning which parts of the context to include in their papers.

Improvement mechanisms: We believe that improvements w.r.t. considering context when conducting SE research would mainly come from paradigm shifts in the SE community and also from changed mindset, awareness and appreciation of SE researchers for research relevance.

---

### 5.1.5 Need for cost-benefit analysis of SE techniques

Four sources called for cost-benefit analysis of SE techniques (S6, S14, S22, S24). S22 argued that "*any modeling or formalizing [in SE research] would have to be lightweight*". S24 recommended "*more research on the economics of RE*". S24 argued that "*lack of concrete knowledge of what organisations can gain from applying state-of-the-art but also time-consuming requirements approaches is one of the major obstacles in getting research results adopted by practitioners*".

---

Synthesis: Although only four sources called for the need for cost-benefit analysis of SE techniques proposed in papers, we think this improvement is among the most important ones. A good example is S6, in which the author argued that: "*it is important to always state where the models [to be used in model-based testing] come from: are they artificial or did they already exist before the experiments*" and that "*one has to argue and evaluate if the time and effort in developing and maintaining such models for a given system does pay off in the end*". S6 clearly criticized the assumption of many papers in the model-based testing literature

---

since the cost of developing such detailed models may be more than the benefits to be gained from model-based testing. In such a scenario, it is almost impossible to "convince" practitioners to adopt such approaches, thus leading to poor relevance.

Improvement mechanisms: Cost-benefit analyses are widely practiced by engineers and managers in the software industry, although relatively few materials are shared in papers or online, e.g., [172]; often since such analyses have sensitive information about companies. Similar to the topics of Value-based software engineering (VBSE) [173] and SE economics [174], there seems to be a need for value-based SE research, i.e., the value (cost-effectiveness) of any proposed SE approach should be assessed. Some recent empirical studies have started to report cost-benefit analyses of the proposed SE approaches, e.g., in [175], cost/benefit drivers of automated testing for a specific project and context were identified and quantitatively measured. Once more and more SE papers conduct quantitatively the cost-benefit analyses of the proposed approaches, practitioners could more objectively assess the applicability of the research.

### 5.1.6 Other improvements involving researchers

Six sources mentioned other improvement ideas for researchers. For example, S18 suggested to "*disseminate work more widely*" (referring to publications in industrial venues).

S19 suggested to "*Keep on moving, i.e., do not stay in your cozy SE corner*". S37 argued that "*SE courses need to present more research. Show off the cool things we invent and discover!*". S54 suggested to "*consider presenting ideas [in SE papers] as patterns and anti-patterns*".

### 5.2 Changes in the academic system

Eight sources mentioned improvement suggestions related to the global academic system. For example, S3 advocated for "*real world benefits and customer focus over h-indices and acceptance rates*". S6 and S9 advocated for "*better reward systems*" in academia.

S8 believed that academia "*should give more freedom of exploration to younger researchers*". S38 suggested that "*hiring and promotion committees must emphasize and reward evidence of impact on engineering practice*". S29 argued along the same lines. S48 suggested that academia should provide resources to support industry collaborations.

Synthesis: Most sources discussed certain (cultural) changes to academia, so that researchers feel encouraged to conduct relevant research.

Improvement mechanisms: To make improvement in this category, changes to the academic culture must be made, which go well beyond just the SE community. The importance of the number of papers and citations has almost been "engraved" in academic culture, and the dream of changing the academic system to put more emphasis on "*real world benefits*" over number of papers and citations may not be easy to realize. Although some (small) progress seems to have been made in some countries. For example, the national Research Excellence Framework (REF) in the UK includes a component on research "impact", which plays a factor in ranking of research institutions.

We should mention and acknowledge, in this outset, that changes (improvements) in the academic system is not (always) *easy* or *simple*. For example, a paper in the field of psychology [88] which investigated research rigor and relevance, identified the stakeholder demands (duties, "pressures") on academic researchers and practitioners, as visualized in Figure 13. For example, that study [88] listed several academic demands/duties for academic researchers which consume their time and energy and discourage them from conducting impactful or industrially-relevant research, e.g., funding council demands, institutional demands (e.g., teaching, service, funding cuts, etc.), audit demands (teaching and research), and professional demands (e.g., committee work, journal editorial boards, etc.).

On the other hand, the study [88] listed several main demands/duties for practitioners, including: business demands (e.g., profitability, viability, future survival and growth), employee demands (e.g., job role, career and personal growth), professional demands (e.g., best practices), and client demands (e.g., usability, credibility, etc.). All the above demands/duties in the either side reduce the possibility of and motivations for joint industry-academia collaborations, which often directly help increase practical relevance of research.

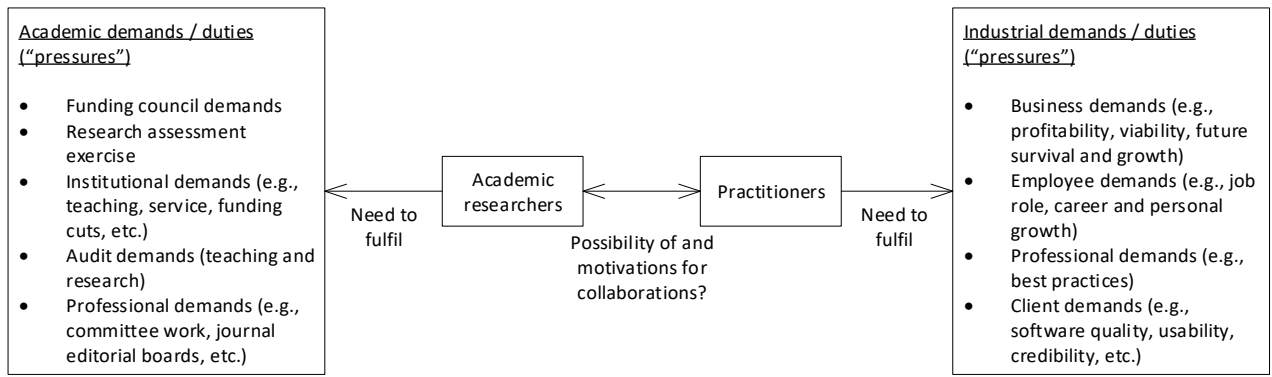

**Figure 13- Stakeholder demands (duties, "pressures") on academic researchers and practitioners. Idea is adopted from: [88]**

### 5.3 Changes in the SE community

Eight sources mentioned improvement suggestions under this category. While some improvement suggestions were about the academic system, some were specifically about the SE community.

S5 suggested "*putting 'impact on practice' as a publication requirement in conferences and journals*". S27 suggested that "*top journals and conferences must acknowledge that context-driven research is needed, valuable, and challenging*". S28 also suggested to "*value research on real problems*". S9 suggested that "*we should stop seeing ourselves as computer scientists*". Along the same lines, S35 suggested "*moving the research tradition from computer science*". S16 suggested that "*academic journals [should] require submissions to note whether the authors have conducted applicability checks*". S41 called for "*Making experience/application tracks first class citizens in SE conferences*", and "*need for more papers focused on understanding the problems*".

> Synthesis: The sources in this category mainly suggested changes in the SE community that would encourage more SE researchers to conduct relevant research. But, as for Section 5.2, most of those changes again involve "cultural" and mindset changes.
>
> Improvement mechanisms: Many of the improvements in this category seems to be "easier" to achieve than those in the "academic system" category, since the SE community is a much smaller community, relatively, compared to the world-wide academic system. Some recent improvements have been made, e.g., establishment of "industrial" tracks in several SE conferences. If more and more researchers embrace this change and conduct more relevant research, some form of "snowballing" effect may occur, and more researchers may move towards conducting more relevant research. As Mahatma Gandhi said: "*Be the change you want to see in the world*".

### 5.4 Changes to the funding system

Five sources mentioned improvement suggestions under this category. S6 mentioned that funding agencies play a major role in the lack of research relevance and impact on industry. It suggested that "*If funding projects changed their metrics and allowed more money and time to be dedicated at developing proper, and usable tools, ... many researchers would do so. In the current environment, the money/time is only enough for a prototype at best*".

S29 suggested that "*funding agencies must help promote and reward collaborative research between public research and industry, thus enabling context-driven research*". S38 made a similar suggestion. S39 suggested that "*research funding mechanisms should integrate the practical needs of SE research*".

> Synthesis: Due to the above fundamental short-comings in many national / international funding systems, it is challenging to conduct truly relevant research. Although there are positive policies in the context of certain funding agencies which encourage, or even require,98 IAC for submitting research grants, e.g., the Horizon 2020 funding system of the European Union, the "Engage" and Collaborative Research and Development (CRD) grants in Canada by the Natural Sciences and Engineering Research Council of Canada (NSERC).

<u>Improvement mechanisms:</u> Changing funding policies and funding systems is rather difficult and often can only be achieved over long-term via lobbying and other influences.

## 6 SUMMARY AND RECOMMENDATIONS

We present a summary of our results, provide recommendations and then discuss the potential threats to validity of the MLR.

### 6.1 Summary

This MLR has synthesized the existing literature and discussions in the SE community about research relevance. Rather than relying on and considering the opinions mentioned in only one or few of the sources, written by one or a few authors, the MLR has collected and synthesized the opinions from *all* the 54 sources about this issue and thus, we believe, it provides a more *holistic* view on the two issues under investigation: root causes of low relevance and suggestions for improving relevance.

As Figure 9 shows, the top-three root causes of low relevance, as discussed in the community, are: (1) Simplistic view (wrong assumptions) about SE in practice (discussed in 13 of the 54 sources); (2) Lack of connection with (experience) industry (discussed in 13 sources); and (3) Wrong identification of research problems (12 sources). We also found that each root cause was mentioned in multiple sources, thus we see that many members of the community share similar opinions on the issue.

As Figure 10 shows, the top-three suggestions for improving research relevance are: (1) Using appropriate research approaches (discussed in 19 sources), such as action-research; (2) Choosing relevant (practical) problems (19 sources); and (3) Collaborating with industry (17 sources). In general, as expected, most of the suggestions for improving research relevance, discussed in the sources, corresponded to the root causes and are, in fact, suggested solutions to address those root causes. For example, S41 suggested adopting the "industry-as-laboratory" approach to replace the "Research-then-transfer" approach, which often lowers relevance.

In Sections 4 and 5, we explained each of the root causes and improvement suggestions by synthesizing the discussions and examples as reported in the sources. Furthermore, for improvement suggestions, we highlighted various improvement mechanisms, when possible, from the sources and from other studies in the SE literature.

We also noticed that, due to the complex nature of the subject, there could be interdependencies among different root causes and improvement suggestions, e.g., "lack of connection with industry" could lead to "wrong identification of research problems" and also "ignoring cost-benefit of applying a SE technique". We clarified that we were aware of this fact during our data extraction and synthesis, and, to minimize any bias in our data synthesis, we decided to record and synthesize the root causes, as they were explicitly reported in the sources (see the example in Figure 8). To ensure that during qualitative analysis (coding) and synthesis, we stayed as close as possible to what the original authors had intended.

Based on the reviewed literature, we believe that the root causes of low relevance and suggestions for improving relevance should be assessed in the context of different actors and scopes, e.g., issues specific to researchers and their mindsets, issues in the (SE) research community, and issues in the academic system.

### 6.2 Recommendations and insights

As discussed in Section 3.9, in addition to the synthesis carried out for each of the root causes and improvement suggestions, we planned a second level of synthesis to aggregate all root causes and improvement suggestions with the goal of providing "actionable" recommendations and insights.

We classify the recommendations by the same actor types, as classified in Sections 4 and 5, as follows.

**Recommendations (improvement suggestions) to individual researchers:**

Based on the reviewed literature, our observations in the community (e.g., in conference debate sessions), and by talking to colleagues in person, we believe that everything would start with an SE researcher's awareness and appreciation for research relevance. Once an SE researcher forms the passion for conducting more relevant research, the "improvement suggestions", as discussed in Section 5, could be followed:

- Using "proper" research approaches
- Choosing relevant (practical) problems
- Collaborating with industry
- Paying attention to "context"
- Carrying out cost-benefit analysis of SE techniques

Also, we should note again that since SE researchers are connected to (or more precisely, embedded in) the SE community, the mindset change and awareness for relevant research, specific for an SE researcher, is often impacted by and could impact the group-level and finally community-level opinions and awareness with respect to research relevance. Moreover, research shows that a committed minority can rapidly change the majority opinion in a population when the committed fraction grows beyond 10% [176]. We are not sure if such a phenomenon would happen in a research community (the SE community in our case). In summary, we observe that such "improvements" (mindset shifts) in the context of individual researchers are inherently "intertwined" with group-level and community-level improvements, as discussed next.

**Recommendations (improvement suggestions) to the SE community:**

It makes sense to again look back at the sources which mentioned community-level improvements:

- **S5**: "*putting 'impact on practice' as a publication requirement in conferences and journals*"
- **S27**: "*top journals and conferences must acknowledge that context-driven research is needed, valuable, and challenging*"
- **S28**: "*value research on real problems*"
- **S9**: "*we should stop seeing ourselves as computer scientists*"
- **S35**: "*moving the research tradition from computer science*"
- **S16**: "*academic journals [should] require submissions to note whether the authors have conducted applicability checks*"
- **S41**: "*making experience/application tracks first class citizens in SE conferences*"

It seems that we are dealing with a "feedback loop" here. The more the community as a whole "values" research on real problems, the more researchers are going to embrace it. After all, our projects and papers are almost always reviewed by other SE researchers who work in another university.

**Improvement suggestions to the funding systems:**

Since funding systems are not our audience in this paper, there is little point mentioning "Recommendations" to the funding system. We have therefore expressed the above title as "improvement suggestions".

Changing funding policies and funding systems, to further support IAC and relevant research, is not easy and often can only be done in long-term via lobbying and other forms of influence. However, almost all of our project proposals are reviewed by other SE or CS researchers who work in other universities. Thus, if they value relevant research and assess such proposals positively, there will be more momentum for relevant SE research.

**Improvement suggestions in the academic system:**

Even perhaps harder than improvements in the funding system, improvements in the classical academic system are not easy. As discussed in Section 5.2, number of papers and citations have almost been "carved" as evaluation metrics in academia, and the dream of changing the academic system to put more emphasis on "*real world benefits*" over number of papers and citations is hard to realize.

From another point of view, by being involved in academic committees and decision making, SE researchers can have influences (even minor) on such improvements. Mahatma Gandhi's words are also relevant here: "*Be the change you want to see in the world*".

## 6.3 Potential threats to validity of the MLR

We carefully identified and addressed potential threats to validity at every step of this MLR study. We followed the guidelines for performing SLRs and benefited from our previous experience, e.g., [134, 153, 177-183]. This sub-section discusses the potential threats to validity and the actions we took to minimize or mitigate them.

<u>Internal validity:</u> A systematic approach has been applied to conduct this study, as described in Section 3. To enable the repeatability of this study, we have defined and reported all the steps, including search engines, search terms, and the inclusion/exclusion criteria that we used. Some problems related to internal validity may have occurred due to the limitation of search engines and terms and bias in applying inclusion/exclusion criteria.

To obtain a finite set of primary studies (sources) for review, it is mandatory to limit the number of search engines and terms used. Such a limitation is a threat for the completeness. To decrease the possibly negative impact of this risk, the first author conducted the searches using the search terms and the other two authors checked the pool to minimize risk of missing relevant sources. In addition, we also conducted forward and backward snowballing [154] on the set of papers already in the pool.

To minimize the risk of bias in applying inclusion/exclusion criteria, the first and second authors voted for the papers to be included. The papers which had conflicting votes have been explicitly discussed by the authors. A consensus was reached for all such papers.

There is also a risk that the process of extracting information from the primary studies is biased. We mitigated this by expressing clear instructions for the data extraction, i.e., the qualitative coding in pink, green, and yellow as described in Section 3.8.2.

<u>Construct validity:</u> The suitability of RQs and the approach used for data extraction address the construct validity for this study. The better the RQs and the data extraction approach reflect the theoretical concepts behind this study, the more valid would this study be from the perspective of this type of validity.

We thoroughly examined the related work and extracted RQs based on the current literature and the experience of the authors. The RQs cover our research goals, which have been answered according to the categorization scheme. Moreover, as reported under internal validity, peer reviews were performed to increase the quality of data extracted. Still, the concept of relevance is an intricate theoretical construct, and there is a risk that its interpretation varies both between individuals and over time. Moreover, the conclusions are consistent with observations of several authors in the SE field. The conclusions are only valid for the SE field and their generality is subject to discussion for other related fields.

## 7 CONCLUSIONS, AND FUTURE WORKS

The main conclusion of this paper is that there has been lots of discussions on research relevance in SE since the 1980's. While some recent positive efforts have been made in this direction, e.g., establishing "industrial" tracks in several SE conferences, we as a community still struggle with research relevance. We also provided an overview of the issue of research relevance in other fields, which seems to be very active.

The paper's main thesis is that we, as the SE community, should conduct and report more relevant research. By synthesizing the root causes of low research relevance and improvement suggestions, provided by prolific community members (via the MLR study), we would like to suggest further attention and discussions on this issue, and aim to add to the already-started "momentum" of recent discussions on this subject.

While the issue of research relevance in SE is real and cannot be ignored, we also need to view our arguments with somewhat critical eyes and consider the possible counterarguments. One of the strongest arguments against research relevance has been that researchers cannot (easily) determine beforehand what (topics) will be relevant later. Also industry often does not know what will be relevant in the future. Our vision is not to stop all high-risk research with long-term potential. However, given the nature of SE and its difference compared to other fields (such as chemistry or theoretical CS), SE is more concerned about increasing effectiveness and efficiency of software development and less about new "product" technologies, such as block-chain in CS or new chemical materials in chemistry. We thus believe that certain SE issues and challenges are quite fundamental and would be suitable candidates for relevant research topics, e.g., increasing effectiveness and efficiency of testing, and identifying such topics, especially in close collaboration with industry, would be the winning point to drive a relevant research endeavor.

On the positive side, we interpret the recent activities done in the SE community to improve relevance as a promising move and we hope to see further progress and more collective efforts in this direction, to which this study can help and provide concrete insights. However, we shall be realistic and, as one senior academic (Dirk Riehle) with a long experience of working in industry put it [8], making the changes to increase the relevance of SE research is "*not easy to achieve and it will take strength and stamina in all of us*".

Another important gap, as pointed out by various scholars, is that the discussions on the issue of research relevance are often conducted in an essayistic style [184]. In a management science paper, Bartunek and Rynes [185] reported that the vast majority of publications on the relevance issue consists of normative opinion statements and only a few are empirical or theory-building. Another paper in management science [55] mentioned that: "*the relevance literature in management studies does not represent a finely structured discourse; the various works are spread across the different sub-disciplines of management studies and only loosely connected to each other*" and that "*On the whole, the programmatic relevance literature largely lacks scientific rigor*". A similar issue has been discussed in the IS community: "*It is not enough for senior IS researchers to call for relevance in IS research. We must also call for an empirically-grounded and rigorous understanding of relevance in the first place*" [120]. Thus, similar to other fields, there is a need for empirically-grounded and rigorous studies on the notion of relevance in SE research. While it is beyond the scope of this study to provide a complete or rigorous understanding of relevance in SE, by synthesizing the evidence and experience-based opinions on this topic from the last 34 years (1985-2019), we aimed to do more than scratching the surface. Our ambition is to encourage more empirically-grounded and rigorous studies on the notion of research relevance in SE.

Future work directions could include the followings: (1) conducting an SLR or MLR on the topic of improvement "mechanisms" and also to develop further "realistic / doable" improvement mechanisms, e.g., devising heuristics and guidelines to choose relevant research problems (topics) that are both rigorous and academically challenging; (2) developing rubrics and benchmarks for systematic relevance assessment of papers and research programs; (3) assessing the topics of current SE research towards labeling them as having lower or higher practical relevance; and (4) as discussed in Section 2.2.5, it would be worthwhile to conduct future synthesis of the issue of research relevance in the pool of studies w.r.t. the types and domains of software systems.